\documentclass[aps,prx,twocolumn,floatfix,longbibliography, twocolumn, superscriptaddress]{revtex4-1}

\usepackage[T2A,T1]{fontenc}
\usepackage{amsmath}
\usepackage{amssymb}
\usepackage{graphicx}
\usepackage{multirow}
\usepackage{natbib}
\usepackage{subcaption}
\usepackage[dvipsnames]{xcolor}
\usepackage[utf8]{inputenc}
\usepackage[colorlinks=true,linkcolor=blue,citecolor=red,urlcolor=blue]{hyperref}
\usepackage{ragged2e}
\usepackage{soul}
\usepackage{physics}

\newcommand{\Mat}{\widehat{C}}

\begin{document}%

%
%%%%%%%%%%%%%%%%%%%%%%%%%%%%%%%
%%%% HEADER
%%%%%%%%%%%%%%%%%%%%%%%%%%%%%%%
%

\title{Polariton-polariton coherent coupling in a molecular spin-superconductor chip}

\author{Carolina del Río}
\affiliation{Instituto de Nanociencia y Materiales de Aragón (INMA), CSIC-Universidad de Zaragoza, 50009, Zaragoza, Spain}

\author{Marcos Rubín-Osanz}
\affiliation{Instituto de Nanociencia y Materiales de Aragón (INMA), CSIC-Universidad de Zaragoza, 50009, Zaragoza, Spain}

\author{David Rodriguez}
\affiliation{Centro de Astrobiología (CAB), CSIC-INTA, 28850 Torrejón de Ardoz, Madrid, Spain}

\author{Sebastián Roca-Jerat}
\affiliation{Instituto de Nanociencia y Materiales de Aragón (INMA), CSIC-Universidad de Zaragoza, 50009, Zaragoza, Spain}

\author{María Carmen Pallarés}
\affiliation{Instituto de Nanociencia y Materiales de Aragón (INMA), CSIC-Universidad de Zaragoza, 50009, Zaragoza, Spain}
\affiliation{Laboratorio de Microscopias Avanzadas (LMA), Universidad de Zaragoza, 50018 Zaragoza, Spain}

\author{J. Alejandro de Sousa }
\affiliation{Instituto de Ciencia de Materiales de Barcelona (ICMAB), CSIC, Barcelona, Spain}

\author{Paweł Pakulski}
\affiliation{Jagiellonian University, Faculty of Chemistry, Krakow, Poland}

\author{José Luis García Palacios}
\affiliation{Instituto de Nanociencia y Materiales de Aragón (INMA), CSIC-Universidad de Zaragoza, 50009, Zaragoza, Spain}

\author{Daniel Granados}
\affiliation{IMDEA-Nanoscience, Cantoblanco, 28049 Madrid, Spain}

\author{Dawid Pinkowicz}
\affiliation{Jagiellonian University, Faculty of Chemistry, Krakow, Poland}

\author{Núria Crivillers}
\affiliation{Instituto de Ciencia de Materiales de Barcelona (ICMAB), CSIC, Barcelona, Spain}

\author{Anabel Lostao}
\affiliation{Instituto de Nanociencia y Materiales de Aragón (INMA), CSIC-Universidad de Zaragoza, 50009, Zaragoza, Spain}
\affiliation{Laboratorio de Microscopias Avanzadas (LMA), Universidad de Zaragoza, 50018 Zaragoza, Spain}
\affiliation{Fundación ARAID, 50018 Zaragoza, Spain}

\author{David Zueco}
\affiliation{Instituto de Nanociencia y Materiales de Aragón (INMA), CSIC-Universidad de Zaragoza, 50009, Zaragoza, Spain}

\author{Alicia Gomez}
\affiliation{Centro de Astrobiología (CAB), CSIC-INTA, 28850 Torrejón de Ardoz, Madrid, Spain}

\author{Fernando Luis}
\affiliation{Instituto de Nanociencia y Materiales de Aragón (INMA), CSIC-Universidad de Zaragoza, 50009, Zaragoza, Spain}
\email{fluis@unizar.es}

\begin{abstract}

The ability to establish coherent communication channels is key for scaling 
up quantum devices. Here, we engineer interactions between distant polaritons, hybrid spin-photon excitations 
formed at different lumped-element superconducting resonators within a chip. The chip consists of several 
resonator pairs, slightly detuned in frequency to make them addressable, capacitively coupled 
within each pair and inductively coupled to a common readout line. They 
interact locally with samples of PTMr and Tripak$^{-}$ organic free radicals, deposited 
onto their inductors, which provide model $S = 1/2$, $g \simeq 2$ spin ensembles. 
Frequency-dependent microwave transmission experiments, performed at very low temperatures, measure polariton 
frequencies as a function of magnetic field in different scenarios. When only one resonator 
within a pair hosts a molecular sample, the results evidence that spins couple remotely
to the empty LER as well as to the local cavity mode. If both resonators interact 
with a spin ensemble, the magnetic field tunes the polariton frequencies relative to each other, on 
account of the different spin-photon interactions at each LER. When polaritons are brought into 
mutual resonance, an avoided level crossing emerges that gives direct spectroscopic 
evidence for a coherent polariton-polariton interaction mediated by the circuit. Pump-probe 
experiments reveal that the excitation of a polariton within a connected pair is felt, thus it can be
read out, by the other one. These observations, backed by model calculations, 
illustrate the control and detection of distant photon-photon and spin-spin correlations and 
entanglement in a scalable modular chip. 

\end{abstract}

\maketitle

%%%%%%%%%%%%%%%%%%%%%%%%%%%%%%%
%%%% INTRO 
%%%%%%%%%%%%%%%%%%%%%%%%%%%%%%%
\section{Introduction}\label{sec:introduction}

Circuit quantum electrodynamics (circuit QED) studies quantum matter interacting with 
photon modes of superconducting cavities \cite{Blais2004,Blais2021}. Besides the 
fundamental interest of accessing strong \cite{Wallraff2004} and ultra-strong 
\cite{Niemczyk2010} light-matter interaction regimes, circuit QED provides also a platform for 
reading out solid-state qubits 
\cite{Wallraff2005}, for generating quantum states of light \cite{Houck2007} and for 
quantum sensing \cite{Bonizzoni2024}, among others. In addition, photons introduce effective 
interactions between qubits coupled to the 
same cavity, which can be exploited for establishing coherent 
communication channels and, therefore, scaling up quantum computing architectures. Such a 
quantum bus has been realized with superconducting qubits \cite{Majer2007} and with spin 
qubits in semiconductors \cite{Borjans2020,Harvey-Collard2022,Dijkema2024}, both of which 
couple to the electric photon field. 

\begin{figure}[htb!]
\centering
\includegraphics[width=0.98\linewidth]{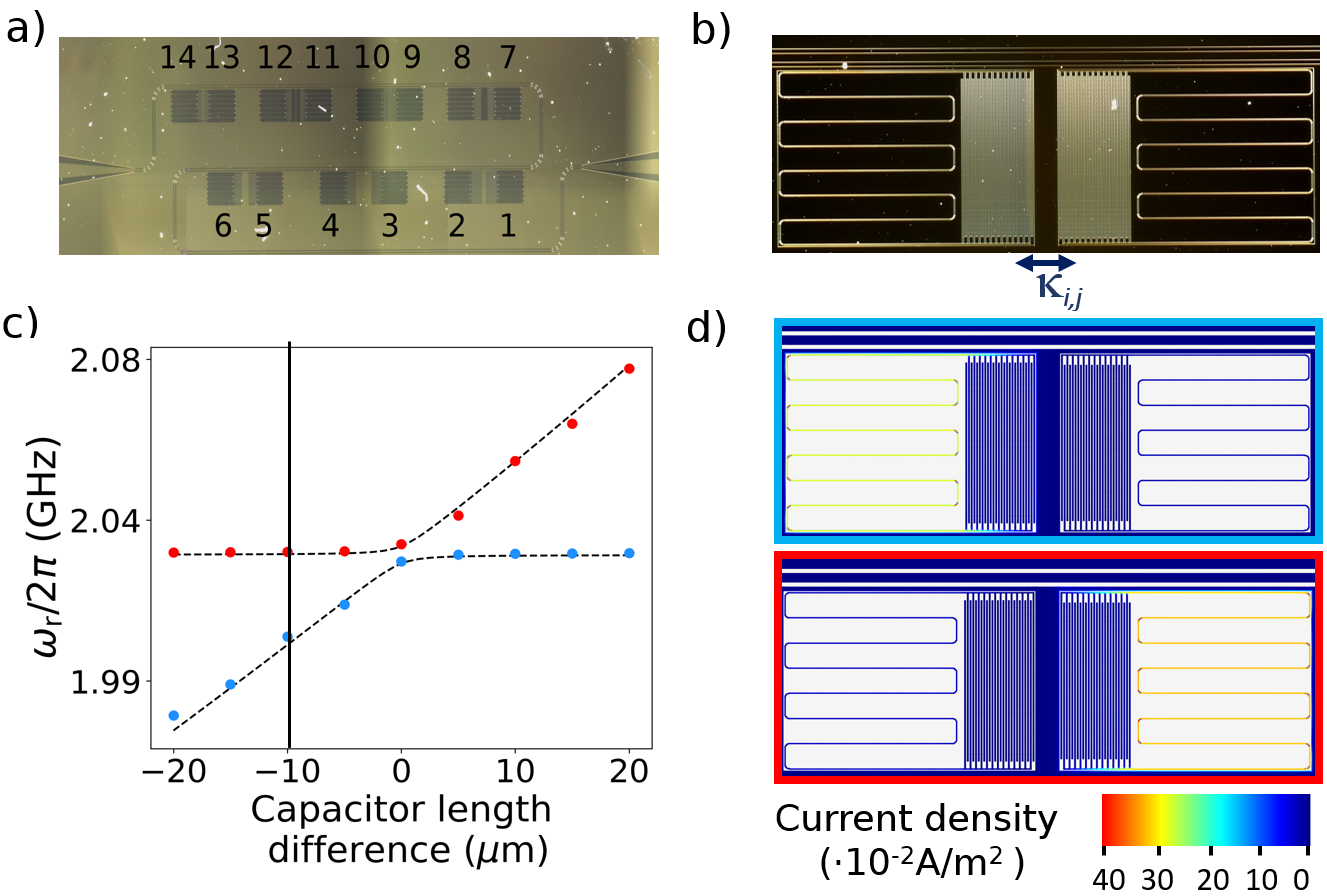}
\captionsetup{justification=justified}
\caption{\justifying a) Image of a chip hosting seven superconducting lumped element resonator (LER) 
pairs coupled to a common transmission line. b) Optical microscopy image of one of these capacitively 
coupled LER pairs. c) Resonance frequencies of a LER pair calculated as a function of the 
difference between their finger capacitor lengths. The coupling introduces a mode 
splitting $\Delta \omega_{\rm r} = 2 \kappa_{i,j}$ when the two LERs 
become identical. d) Single-photon current density simulations for the two resonant modes of a 
detuned LER pair (vertical line in panel c). Further details on the 
design, simulation, fabrication and characterization of the chips are given in Appendices 
\ref{ssec:design} and \ref{ssec:microwave}.}
\label{fig:set-up}
\end{figure}

The magnetic spin-photon interaction is much weaker, 
but it is enhanced in the case of macroscopic spin ensembles 
\cite{Dicke1954,Schuster2010,Kubo2010,Wu2010,Amsuss2011,Ghirri2015,Bonizzoni2017,Mergenthaler2017,Lenz2021,Rollano2022}. 
This has allowed coupling collective excitations of two distant 
samples, either organic free radicals \cite{Ghirri2016} or NV$^{-}$ centers 
in diamond crystals \cite{Astner2017}, through a coplanar superconducting resonator. 
The need of coupling both ensembles to the same cavity mode adds, however, 
some constrains. For instance, they must be located at specific positions determined 
by the photon magnetic field profile. Besides, one of the resulting coupled states is 
always dark, which precludes a direct determination of the circuit mediated interaction by 
spectroscopic measurements.

In this work, we move beyond these limitations and explore a modular hybrid quantum platform, 
in which distant polaritons, i.e. light-matter hybrid states generated by local spin-photon 
interactions at different lumped-element superconducting cavities, can be tuned by a magnetic field, 
addressed in situ in a multiplexable manner and coherently coupled. This scheme offers the 
possibility of generating, tuning and detecting excitations that simultaneously involve 
spin {\em and} photon entanglement. It has also a high design flexibility, e.g. to build in 
asymmetries that turn each polariton distinguishable from the others, and it is inherently 
scalable to multiple nodes, which would make it relevant for applications to 
quantum simulation \cite{Chiesa2015,Roca2025simulating} and quantum computing \cite{Jenkins2016,Chiesa2023}.

\section{Experimental setup}\label{sec:setup}

\subsection{Superconducting circuits}\label{ssec:circuit}

The device used in this work is sketched in Figs. \ref{fig:set-up}a 
and \ref{fig:set-up}b, and described in detail in Appendix~\ref{ssec:design}. The chip consists of 
$7$ pairs of superconducting lumped-element $LC$ resonators (LERs), all coupled to 
the same readout  line \cite{Rollano2022,Rubin2024}. They are fabricated by maskless optical 
lithography on a $125$~nm thick NbTiN film deposited by sputtering onto a silicon wafer. The 
relevant properties of each superconducting LER, i.e., its resonance frequency 
$\omega_{\mathrm{r},i}$ and the photon loss rate $\kappa_{i}$ ($i = 1,\ldots,14$), can be tailored 
within a wide range without affecting the transmission properties of the device. In the chip of 
Fig.~\ref{fig:set-up}a, different $\omega_{\mathrm{r},i}/2 \pi$ ranging 
from $1.7$~GHz to $3.3$~GHz were tuned by varying the capacitor geometries while leaving all inductors, which 
provide the coupling to the spin samples, the same. This allows multiplexing the LER readout, as 
shown in Appendices \ref{ssec:design} and \ref{ssec:microwave}. The two LERs 
within each pair are capacitively coupled, with a coupling 
strength $\kappa_{i,j}$ that is controlled by the separation between their respective 
capacitors. If they were identical, this coupling would lead to 
an avoided level crossing, shown in 
Fig.~\ref{fig:set-up}c,  with a frequency splitting $\Delta \omega_{\rm r} = 2 \kappa_{i,j}$ 
\cite{Noroozian2012} between symmetric and anti-symmetric modes delocalized all over the 
LER pair (Fig.~\ref{fig:Sonnet simulations} from Appendix~\ref{ssec:design}). Here, we detune 
the `bare' resonance frequencies $\omega_{\mathrm{r},i}$ and 
$\omega_{\mathrm{r},j}$ of the two LERs by making the lengths of their interdigitated capacitors 
different. When $|\omega_{\mathrm{r},i}-\omega_{\mathrm{r},j}| > \kappa_{i,j}$, the photon 
magnetic field becomes predominantly localized at either one or the other 
LER (Fig.~\ref{fig:set-up}d 
and Fig.~\ref{fig:Sonnet simulations}). This sort of `dispersive' condition ensures that, in 
spite of their mutual coupling, each of them can still be addressed by choosing the appropriate 
frequency.

\begin{figure}[htb!]
    \centering
    \includegraphics[width=1.01\linewidth]{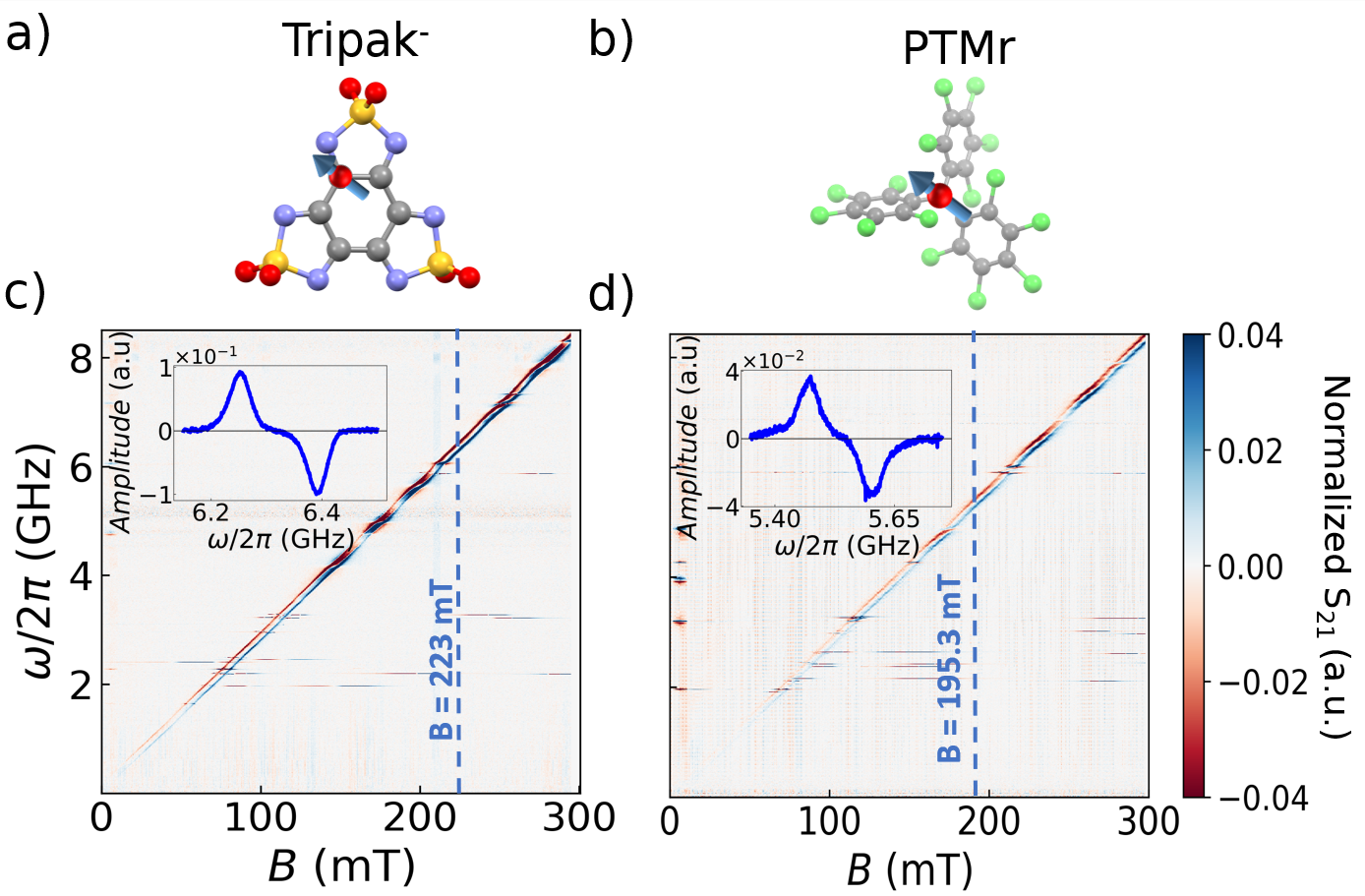}
    \caption{ \justifying  Molecular structures of the organic free radical molecules used in this study: a) Tripak$^{-}$ 
    \cite{Pakulski2024}; b) PTMr \cite{Schafter2023}. Here, carbon atoms are represented in grey, chlorine  atoms in green, 
    nitrogen atoms in blue, sulfur atoms in yellow and oxygen atoms in red. c) and d) Broadband microwave absorption spectra 
    of Tripak$^{-}$ and PTMr, respectively, measured at $T = 11$~mK by coupling them to a superconducting transmission line. 
    The insets show normalized transmission data measured at the magnetic fields indicated by vertical dashed lines.}
    \label{fig:spin-system}
\end{figure}

\subsection{Synthesis, characterization and integration of model spin ensembles}\label{ssec:spins}
These LERs host samples of two different organic free-radical molecules. The 
molecular structures of perchlorotriphenylmethyl (PTMr) and 
benzo[1,2-c:3,4-c':5,6-c'']tris([1,2,5]thiadiazole)2,2,5,5,8,8-hexaoxide (Tripak$^{-}$) 
radicals, synthesized by following chemical 
procedures reported elsewhere \cite{Armet1987,Schafter2023,Pakulski2024}, are shown in 
Figs. \ref{fig:spin-system}a and \ref{fig:spin-system}b. These molecules were 
transferred from solution onto the LER inductors and then left dry in the dark 
(see Appendix~\ref{ssec:integration} for details). The solution contained also a polystyrene 
polymer that prevents agglomeration of the radical molecules in the dry deposits. 
Each of these deposits is then highly stable at ambient conditions and provides a model ensemble of 
$N\sim 5\times10^{12}-5\times10^{14}$ identical $S=1/2$ spins 
with a quasi-isotropic $g$ factor, very close to that of a single electron. The latter property 
leads, at any magnetic field $B$ and despite the random molecular orientations, to a 
well-defined narrow spin resonance, which was characterized as a function of $B$ by coupling these 
spin samples to a superconducting waveguide. Representative results, measured 
at very low temperatures and normalized as described in \cite{Gimeno2023,Roca2025}, are shown 
in Figs. \ref{fig:spin-system}c and \ref{fig:spin-system}d. Additional data are shown in 
Fig.~\ref{fig:Espectros} from Appendix~\ref{ssec:integration}. The spin resonance frequency 
$\Omega_{S}$ increases linearly with $B$, as expected for a paramagnetic system. 
The slightly different slopes reveal different $g$ factors, $2.001$ for PTMr and 
$2.003$ for Tripak$^{-}$. The resonance line shape, shown in the insets of 
Fig.~\ref{fig:spin-system}, allows estimating the spin line width $\gamma/2 \pi \simeq 7-12$~MHz for both samples, 
depending on concentration.

\begin{figure}[htb!]
 \centering
\includegraphics[width=1.0\linewidth]{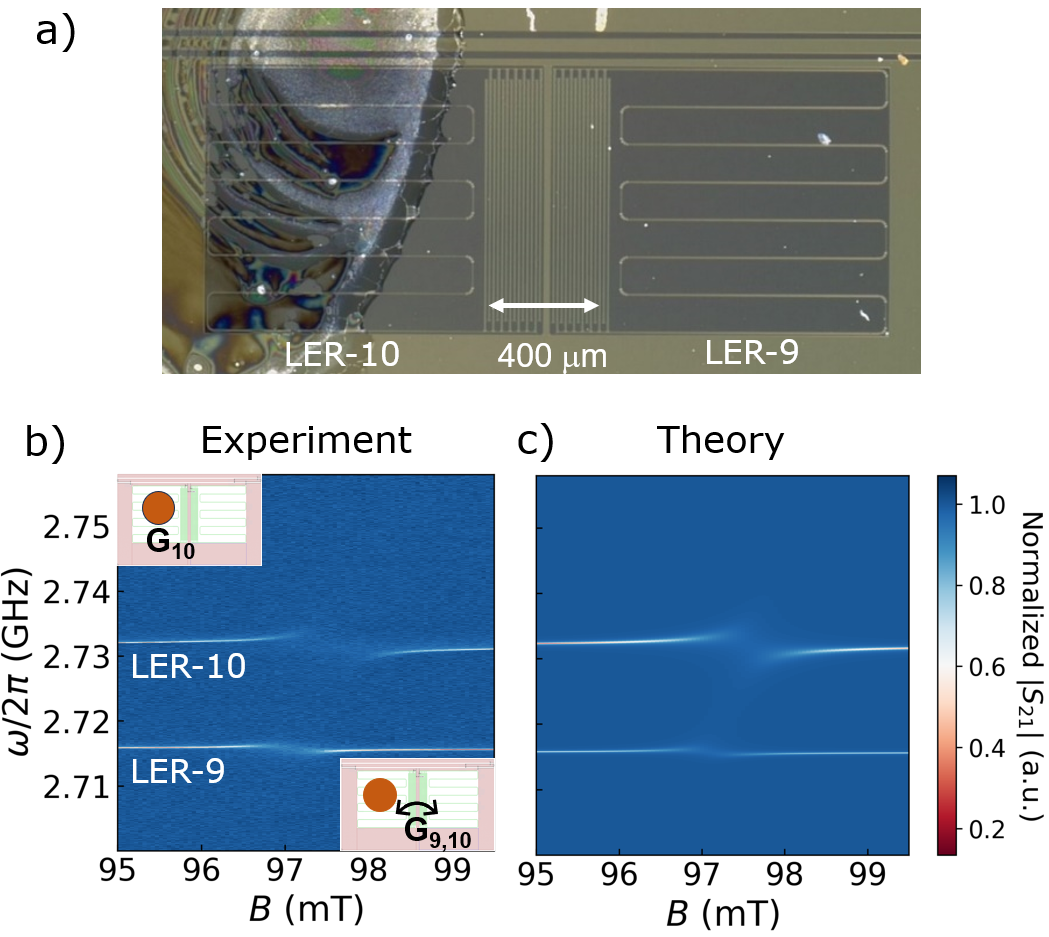}
 \caption{ \justifying a) Optical microscopy image of a $N\sim 5 \times 10^{12}$ PTMr free 
 radical deposit on the inductor of LER-10, forming a 
 coupled pair with LER-9, which remains empty. b) Colour plot of the microwave 
 transmission amplitude $\vert S_{21} \vert$ measured, at $T=11$~mK, near the resonance 
 frequencies of the two LERs. The data show a detectable coupling of the spins to both 
 resonant modes, being $G_{10}/2 \pi = 5.4$~MHz 
 the coupling strength to the `local' LER-10 and $G_{9,10}/2 \pi = 2.5$~MHz the 
 remote interaction with LER-9. c) Theoretical calculation of the microwave transmission 
 obtained from Eq.~\eqref{eq:hamiltonian} by using input-output theory 
 (Appendices~\ref{ssec:input-output} and \ref{ssec:application}),
 for $G_{10}/2 \pi=5.4$~MHz and $\kappa_{9,10}/2 \pi = 6.5$~MHz.}
 \label{fig:remote-spin}
\end{figure}

\section{Results}\label{sec:results}

\subsection{One spin ensemble coupled to a LER pair: remote spin photon coupling}\label{ssec:1spin}
Let's now move into the response of LER-spin units. We first consider the situation, 
shown in Fig.~\ref{fig:remote-spin}a, in which only one of the LERs, LER-10, hosts a sample of 
$N \sim 5 \times 10^{12}$ PTMr molecules, while the second one, LER-9, remains empty. 
With this setup, we aim to compare the couplings of the same spin ensemble to 
local and remote photon modes. Microwave transmission 
data $S_{21}$ were measured at very low temperature as a function of frequency and 
magnetic field using the setup described in Appendices~\ref{ssec:cryogenic} and 
\ref{ssec:microwave}. Results are shown in Fig.~\ref{fig:remote-spin}b. The two modes appear as horizontal lines at slightly different 
frequencies. When the magnetic field brings the spins to resonance with LER-10, the 
transmission shows a frequency shift, a broadening and a reduced visibility of this mode, which characterize
the local spin-photon coupling. Fitting these data using input-output theory (described in 
Appendices~\ref{ssec:input-output} and \ref{ssec:application}) gives a 
collective spin-photon coupling strength $G_{10}/2 \pi = 5.4$~MHz, which corresponds to high 
cooperativity $C \equiv G_{10}^{2}/\kappa \gamma \sim 70$. Interestingly, the same spin ensemble also couples 
to the photon mode of the nominally empty resonator LER-9. The remote coupling rate 
$G_{9,10}/2 \pi \simeq 2.5$~MHz, is only two times smaller than the local $G_{10}$. 
Hybrid excitations generated at LER-10 are therefore detected by measuring 
the response of LER-9. 

These data can be understood within the framework of the `two-resonator' circuit QED 
theory \cite{Mariantoni2008}, which generalizes the usual Jaynes-Cummings formalism to the case 
of multiple cavities. The Hamiltonian describing the hybrid system can be written as follows 

\begin{eqnarray}
    \label{eq:hamiltonian}
    \mathcal{H} = &&\sum_{q = i,j} \left( \omega_{\mathrm{r},q}a_q^\dagger a_q + \Omega_{S,q}S_q^z \right) \\ \nonumber
    &+& \sum_{q=i,j} G_q\left(a_qS_q^+ + a_q^\dagger S_q^-\right) \\ \nonumber
    &+& \kappa_{i,j}\left(a_i^\dagger a_j + a_ia_j^\dagger\right)\ ,
\end{eqnarray}

\noindent where $a_q$ and $a_q^{\dagger}$ represent the photon annihilation and creation, respectively,  
operators in each cavity $q=i,j$, and $S_q^\alpha$ with $\alpha = \{z,+,-\}$ represent the 
conventional spin-$1/2$ operators. For the system of Fig.~\ref{fig:remote-spin}, $i=9$ and $j=10$, 
and only $G_{10}$ and $\Omega_{10}$ are nonzero since LER-9 is empty. Solving for the eigenvalues and eigenstates 
of \eqref{eq:hamiltonian}, as detailed in 
Appendix~\ref{sssec:1-spin-theory}, shows that the remote $G_{9,10}$ is a direct manifestation of the 
interactions brought about by the circuit and gives the relation 

\begin{equation}
G_{9,10} = \frac{G_{10}}{\sqrt{2}} \sqrt{1-\frac{\vert \omega_{\mathrm{r},9}-\omega_{\mathrm{r},10} \vert}
{\sqrt{\vert \omega_{\mathrm{r},9}-\omega_{\mathrm{r},10} \vert^{2} + 4 \kappa_{9,10}^{2}}}}\ .
\label{eq:G12-1}
\end{equation}

\noindent Equation \eqref{eq:G12-1} allows estimating $\kappa_{9,10} = 6.5$~MHz, which is reasonably close to 
the value $\kappa_{9,10} = 5.8$~MHz derived 
from the circuit simulation (Appendix \ref{ssec:cryogenic}). The application of input-output 
theory to this model (see Appendix~\ref{sssec:1-spin-application}), provides also a good 
description of the transmission data, as shown by Fig.~\ref{fig:remote-spin}c, and 
confirms that LER-9 does not directly couple to the spins but, rather, to the hybrid LER-10 
polariton.

\begin{figure}[htb!]
\centering
\includegraphics[width=1.01\linewidth]{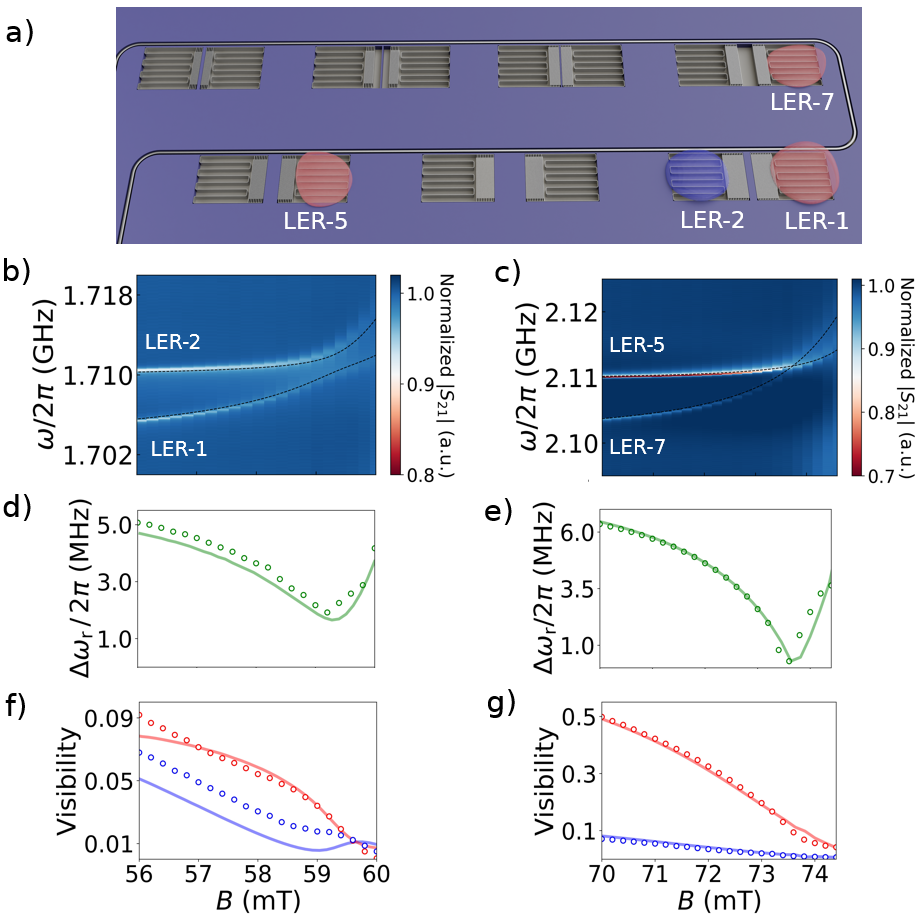}
\caption{ \justifying Experimental setup for polariton-polariton coupling. Two LERs belonging to either a coupled pair 
(LER-1 and LER-2) or to two different and distant pairs (LER-5 and LER-7) host different molecular spin ensembles 
(blue: Tripak$^{-}$, orange: PTMr). 
Panels b), d), f) and c), e), g) provide experimental data and simulations obtained for the coupled and the uncoupled 
LERs, respectively: colour plots of the microwave transmission measured at $T=11$~mK near the resonances of each spin 
ensemble with its local LER (b, c); experimental (dots) and theoretical (lines) frequency differences between the 
two upper polaritons as a function 
of magnetic field (d, e); experimental (dots) and theoretical (lines) visibilities of these modes as a function of 
magnetic field (f, g). The fits are based on the model outlined in the text 
[Eq.~\eqref{eq:hamiltonian}] and detailed in Appendix~\ref{sssec:2-spin-theory}, with parameters 
$G_{1}/2 \pi = 19.5$~MHz, $G_{2}/2 \pi = 8.5$~MHz and $\kappa_{1,2}/2 \pi = 1.06$~MHz for the 
coupled LER pair and $G_{7}/2 \pi = 22$~MHz, $G_{5}/2 \pi = 9.7$~MHz and 
$\kappa_{5,7}/2 \pi = 0$ for the uncoupled LERs. }
\label{fig:spin-spin}
\end{figure}

\subsection{Two spin ensembles coupled to a LER pair:polariton-polariton coupling}\label{ssec:2spin}
Our second study case is illustrated by Fig.~\ref{fig:spin-spin}a 
(see also Fig.~\ref{fig:autoval_LERs1and2} in Appendix \ref{sssec:2-spin-application}). 
Here, LER-1 ($\omega_{\mathrm{r},1}/2 \pi=1.703$~GHz ) and LER-2 
($\omega_{\mathrm{r},2}/2 \pi=1.710$~GHz) within a pair host 
approximately $5\times10^{14}$ PTMr and Tripak$^{-}$ radical molecules, 
respectively. This setup aims to couple distant polaritons via the circuit. 
For comparison purposes, we have also performed the same study on two different PTMr samples 
placed onto further distant $\simeq 2.1$~GHz LER-5 and LER-7 belonging to different 
pairs (Fig.~\ref{fig:spin-spin}a). This allows comparing experimentally the response of a coupled 
LER pair to that of two uncoupled LERs. Microwave transmission data measured near the 
spin resonances $\Omega_{S,i} = \omega_{{\rm r},i}$ are shown in 
Figs. \ref{fig:spin-spin}b and \ref{fig:spin-spin}c, while results obtained 
over wider frequency and magnetic field regions are given in Appendices~\ref{sec:data} 
and \ref{sssec:2-spin-application}.  

\begin{figure}[htb!]
\centering
\includegraphics[width=0.85\columnwidth]{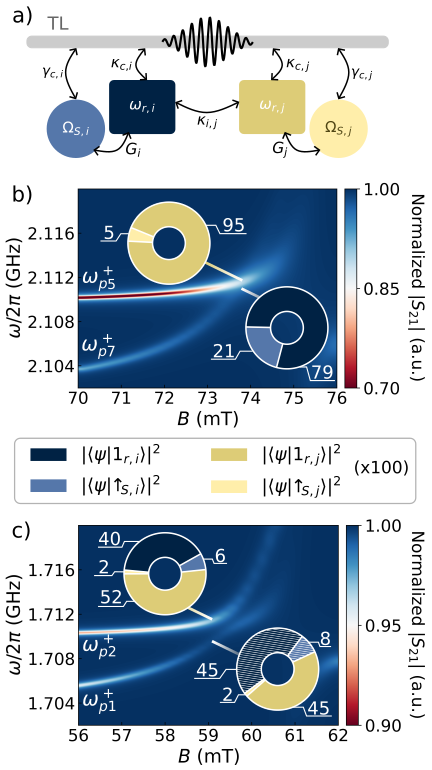}
\caption{\justifying a) Sketch of the polariton-polariton coupling model showing the main 
interactions present in the system \eqref{eq:hamiltonian}. Details of the eigenstates and of 
the microwave transmission calculations are described in Appendix~\ref{sec:theory}. b) and c) 
Simulated $2D$ microwave transmission plots for two uncoupled ($\kappa_{i,j}=0$) and two coupled 
($\kappa_{i,j} = 1.06$~MHz) LERs, respectively. The insets show the probabilities of 
the single spin and single photon excitation states in each polariton. Striped sectors in the 
inset of panel c) signal negative probability amplitudes.}
\label{fig:model}
\end{figure}

The local spin-photon couplings reach high cooperativity for LER-2 and LER-5 ($G_{2}/2 \pi = 8.5$~MHz, 
$G_{5}/2 \pi = 9.7$~MHz) and the strong coupling regime for LER-1 and LER-7 ($G_{1}/2 \pi = 19.5$~MHz 
$G_{7}/2 \pi = 22$~MHz). The different 
coupling strengths and, to a lesser extent, the different spin 
resonance frequencies of PTMr and Tripak$^{-}$ (see Fig.~\ref{fig:spin-system}) allow 
tuning the upper polariton branches of different LERs relative to each other,
eventually bringing them into mutual resonance as shown in Figs. \ref{fig:spin-spin}b and 
\ref{fig:spin-spin}c. In the case of the uncoupled LER-5 and LER-7 
(Fig.~\ref{fig:spin-spin}c and \ref{fig:spin-spin}e), 
the two branches then simply cross. By contrast, for the two coupled LERs 
(Fig.~\ref{fig:spin-spin}b and \ref{fig:spin-spin}d) there is an avoided level crossing with a 
gap $\Delta \omega^{+}_{\rm p}/2 \pi \simeq 1.9$~MHz near $59.2$ mT. 
This behaviour resembles the mode splitting of two identical coupled LERs, shown in 
Fig.~\ref{fig:set-up}c), only here we deal with polaritons and the tuning is achieved in situ 
by the magnetic field, exploiting their different spin components. By analogy, we conclude that 
the gap $\Delta \omega^{+}_{\rm p}$ is a direct manifestation of the polariton-polariton 
interaction that allows estimating the coupling constant 
$J^{+}_{1,2}/2 \pi \simeq \Delta \omega^{+}_{\rm p}/4 \pi  = 0.96$~MHz. 
Furthermore, near the anticrossing 
the visibilities of the two coupled polariton branches, shown in Fig.~\ref{fig:spin-spin}f, go 
through a local minimum and a local maximum, respectively; i.e., the coupled modes become then 
"brighter" and "darker", which suggests the formation of delocalized photon states with 
dominantly symmetric and antisymmetric characters. This effect is not observed for the two 
uncoupled LERs (Fig.~\ref{fig:spin-spin}g). \newline

Hamiltonian \eqref{eq:hamiltonian} helps again in getting a better insight into these 
results. In this case, and as it's sketched in Fig.~\ref{fig:model}a, the two LERs interact with a different 
spin sample, with respective coupling strengths $G_{i}$ and $G_{j}$. As 
shown in Figs. \ref{fig:spin-spin}d and \ref{fig:spin-spin}e, this model  
accounts for the onset of a finite gap between the two polariton branches when $\kappa_{i,j} \neq 0$, 
which is given by the following expression:

\begin{equation}
\Delta \omega^{+}_{\rm p} = 2J^{+}_{i,j}=2 \kappa_{i,j} \cos{\frac{\theta_{i}}{2}}\cos{\frac{\theta_{j}}{2}},
\label{eq:Delta-omegap}
\end{equation}

\noindent where $\theta_{i} = \tan^{-1}\left( {2G_{i}/\Delta_{i}} \right)$, 
with $\Delta_{i} = \Omega_{S,i}-\omega_{\rm{r},i}$ (and equivalently for LER-$j$) measure the 
spin components of the two polaritons. The same model, combined with the input-output 
formalism that includes the couplings of all components to the readout line 
(see Fig.~\ref{fig:model}a and Appendices~\ref{sssec:2-spin-theory}, \ref{ssec:input-output}
and \ref{sssec:2-spin-application} for details) describes 
the overall features observed in the microwave transmission experiments. This is confirmed by 
comparing the simulations in Figs. \ref{fig:model}b and \ref{fig:model}c to the experimental 
results shown in Figs. \ref{fig:spin-spin}c and \ref{fig:spin-spin}b, respectively. The model 
also accounts for the changes in polariton visibilities that take place near the anticrossing 
(Figs. \ref{fig:spin-spin}f and \ref{fig:spin-spin}g). In all these simulations, $\kappa_{i,j}$ is the only free parameter, the rest being obtained 
from independent fits to the data. The estimated $\kappa_{1,2}/2 \pi = 1.06$~MHz turns out to be a factor two smaller 
than $\kappa_{1,2}/2 \pi = 2.4$~MHz obtained from the circuit simulations 
(Annex \ref{ssec:design}). \newline

The simulations also provide information on the nature of the polaritonic states and on how they are 
modified by their mutual interaction and by external parameters. The probabilities of the 
polaritons in the basis of local single-photon and single-spin excitations are shown as insets to 
Figs. \ref{fig:model}b and \ref{fig:model}c. For $\kappa_{i,j} = 0$, spin and photon excitations 
hybridize locally, but there are no correlations between `left' and `right' excitations. By contrast, for 
$\kappa_{i,j} \neq 0$ the polaritons acquire finite amplitudes of spin and photon excitations 
arising from 
both LERs (Fig.~\ref{fig:2spin2ler_theory} from Appendix~\ref{sssec:2-spin-theory}) when they are brought into resonance by the magnetic field. Near the anticrossing, the photonic 
component approaches symmetric and antisymmetric combinations of the local photon modes, 
and the entanglement between these modes becomes then maximum (Fig.~\ref{fig:2s2r_photon_entanglement} in Appendix~\ref{sssec:2-spin-theory}). 
By contrast, and as discussed in Appendix~\ref{sssec:2-spin-theory}, the opposite limit of 
far detuned, nearly localized polaritons, 
leads to a circuit mediated effective transverse interaction $J(S_{i}^{+}S_{j}^{-}+S_{i}^{-}S_{j}^{+})$ between the 
excitations of the two spin ensembles. This then resembles the situation met when two spin ensembles 
couple to just a cavity mode in the dispersive regime \cite{Majer2007,Amsuss2011,Ghirri2016}. 

\begin{figure}[bt!]
\centering
\includegraphics[width=\columnwidth]{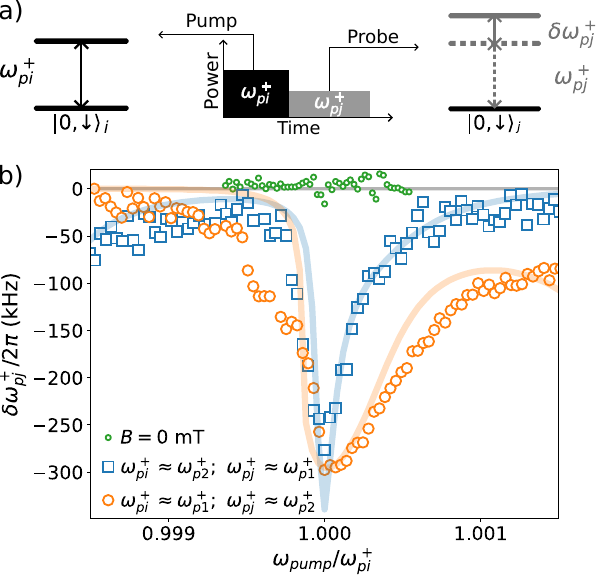}
\caption{\justifying Polariton-polariton pump-probe measurements. a) Sketch of the measurement 
protocol. A short high power microwave pulse resonantly excites one polariton at a frequency 
$\omega^{+}_{\mathrm{p},i}$. Then, a series of lower 
power pulses probe the frequency shift  $\delta \omega^{+}_{\mathrm{p},j}$ generated on the other 
polariton. b) Experimental frequency shift measured at $T=11$~mK by pumping LER-1 polariton 
($\omega^{+}_{\mathrm{p},1}/2 \pi \simeq  1.706$~GHz) 
and detecting changes in the LER-2 polariton (orange circles) and vice versa (blue open 
squares). Data measured at $B=0$, when 
excitations at both LER correspond to pure photon modes and no dispersive shift is observed, are 
also shown as green circles. In these experiments LER-1 was coupled to a PTMr spin ensemble 
whereas LER-2 was coupled to a Tripak$^{-}$ 
sample. The lines are theoretical predictions for the dynamical response 
of the system sketched in Fig.~\ref{fig:model}a obtained by solving the 
time-dependent master equations as described in [\onlinecite{rubin2025}].}
\label{fig:dispersive}
\end{figure}

\subsection{Polariton-polariton dispersive readout}\label{ssec:pump-probe}
The dual character of the polariton excitations and the existence of spin-spin and photon-photon remote correlations 
can be probed by performing pump-probe experiments with microwave pulses that use one of the 
polaritons to sense the excitation of the other one. The protocol and the experimental set-up 
are sketched in Figs. \ref{fig:dispersive} and Appendix~\ref{ssec:pulsed}, respectively. A high-
power short `pump' pulse resonantly excites a polariton on one side of the LER pair. This is 
then followed by a set of `probe' pulses with frequencies scanning that 
of the other polariton that allows detecting any changes in its resonance frequency. \newline

Results of reference experiments performed at zero field, when the LER excitations are pure 
photonic modes, are shown in Figs. \ref{fig:dispersive} and Appendix~\ref{ssec:photon-photon}.  
They show no measurable shift, as expected for two coupled resonators, whose frequencies must 
remain insensitive to changes in their respective populations. By contrast, when $B = 57$ mT, 
near the avoided level crossing in 
Fig.~\ref{fig:spin-spin}b, both polaritons 
acquire a finite spin component (see Fig.~\ref{fig:model}c). In this case, the initial excitation 
of one mode, 
at a frequency $\omega^{+}_{\mathrm{p},i}$ does generate a measurable shift 
$\delta \omega^{+}_{\mathrm{p},j}$ in the 
frequency $\omega^{+}_{\mathrm{p},j}$ of the other one. Examples of the readout process and 
results obtained at 
different magnetic fields are given in Appendix~\ref{ssec:polariton-polariton}.\newline

The measurement of the frequency shift can be exploited to perform polariton spectroscopy. 
For this, $\omega_{\rm pump}$ is swept 
across $\omega^{+}_{{\mathrm{p}},i}$. The ensuing $\delta \omega^{+}_{\mathrm{p},j}$ then maps the distribution of polariton 
excitation frequencies. The results, shown in Fig.~\ref{fig:dispersive}b, agree with 
the polariton line shapes derived from transmission experiments (Fig.~\ref{fig:spin-spin}). 
These measurements also show that the two polaritons can perform both roles, either 
as `spins' or as `probes'. They're also in good agreement with theoretical predictions 
(solid lines in Fig. \ref{fig:dispersive}b) that have been 
obtained by solving the master equations governing the time-dependent response of the 
coupled system, as described in [\onlinecite{rubin2025}]. These results give further insight into 
the upper polariton wave functions. We find polariton line widths $\kappa^{+}_{1}/2 \pi \simeq 0.5$~MHz and 
$\kappa^{+}_{2}/2 \pi \simeq 0.2$~MHz, which lie in between the pure photon and spin line widths, as expected. 
Their values are determined by the ratio $G_{i}/\Delta_{i}$. More strongly coupled spins (in 
this case at LER-1) give then rise to broader polariton excitations. \newline

\section{Conclusions}\label{sec:conclusions}
In summary, we have tested in different scenarios a modular platform whose basic nodes are 
hybrid spin-photon polaritons. The results show that each of these nodes can be addressed and read 
out by quite standard microwave transmission experiments. Besides, they can be deterministically  
coupled via the circuit, with coupling strengths $\sim \Delta \omega_{\mathrm{p}}^{+}/2$ that 
can be made strong enough to overcome polariton losses. This platform opens possibilities to generate and control 
in situ effective interactions and entanglement between distant spins and photons using external parameters 
as `control knobs'. In spite of the fact that the LERs mutual capacitive couplings 
are fixed by the circuit geometry, their effect on the spin and photon polariton components can be tuned by changing the 
magnetic field, and also driving power (Appendix~\ref{sec:power}) or temperature 
(Appendix~\ref{sec:temperature}). This scheme then offers an 
alternative route to realize some key elements of quantum circuits such as 
switches \cite{Mariantoni2008} or filters \cite{Bronn2015}, which are based on 
coupling two or more resonators. In combination with recently developed techniques to enhance the single spin–photon 
coupling \cite{rubin2025}, this work contributes to scaling up hybrid quantum technologies based on molecular 
spins \cite{Jenkins2016,Chiesa2023,Roca2025simulating} by establishing hybrid spin–photon polaritons as addressable nodes in 
superconducting quantum circuits.

%
%%%%%%%%%%%%%%%%%%%%%%%%%%%%%%%
%% ACKNOWLEDGMENTS
%%%%%%%%%%%%%%%%%%%%%%%%%%%%%%%
%

\begin{acknowledgments}
The authors acknowledge useful discussion with M. J. Martínez-Pérez. 
This work has received support from grants TED2021-131447B-C21, TED2021-131447B-C22, PID2022-140923NB-C21, 
PID2022-141393OB-I00, CEX2023-001263-S and CEX2023-001286-S funded by MCIN/AEI/10.13039/501100011033, 
ERDF `A way of making Europe' and ESF 
`Investing in your future', from the Gobierno de Arag\'on grant E09-23R-QMAD, from the European Union 
Horizon 2020 research and innovation programme through FET-
OPEN grant FATMOLS-No862893, from the Novo Nordisk Foundation Exploratory Interdisciplinary Synergy 
Programme 2021 through grant NNF21OC0070832, from QUANTERA project OpTriBits (AEI: PC2024-153480), 
and from the Spanish Ministry for Digital Transformation and Civil 
Service and Next Generation EU through the Quantum Spain project (Digital Spain 2026 Agenda). It also 
forms part of the Advanced Materials and Quantum Communication programmes with 
funding from European Union Next Generation EU (PRTR-C17.I1), MCIN, Gobierno de 
Arag\'on, and CSIC (QTEP-PT-001). D.P. acknowledges the financial support of the National 
Science Centre Poland within the Opus project 2020/37/B/ST5/02735
\end{acknowledgments}

\appendix

\section{Experimental details}\label{sec:expdetails}

\subsection{Circuit design and fabrication}\label{ssec:design}

\begin{figure*}
\includegraphics[width=0.95\linewidth, angle=0]{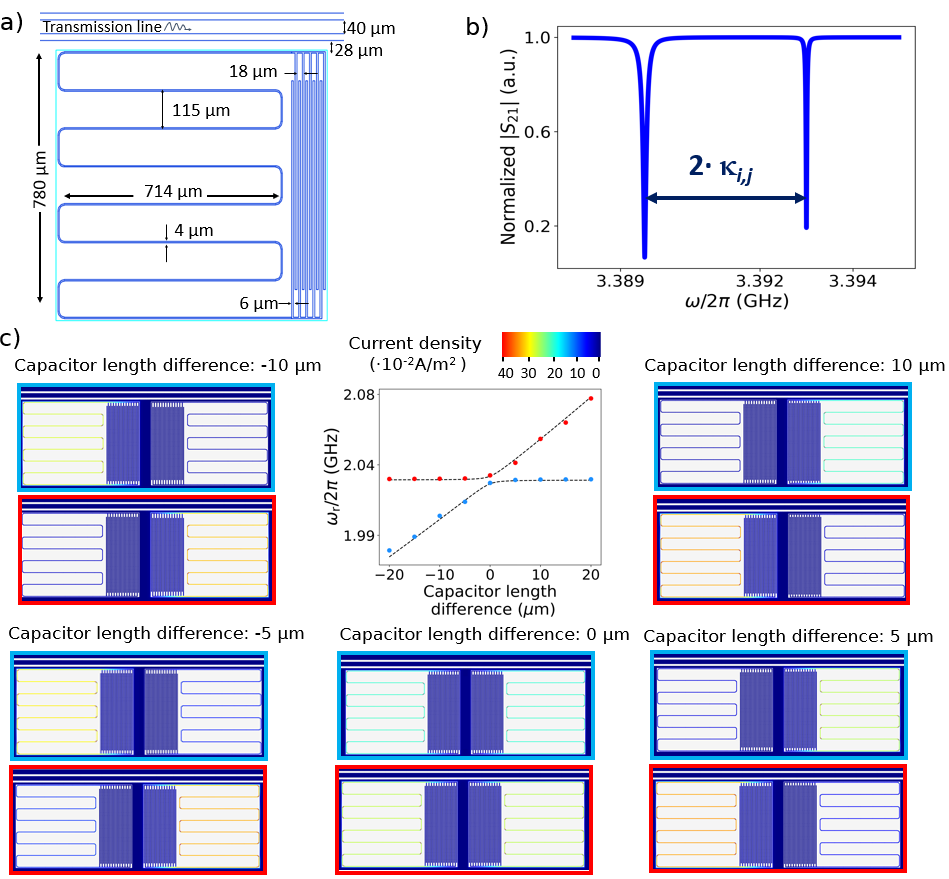}
\caption{\justifying a) Sketch of a superconducting LER showing the $LC$ circuit dimensions. b) Sonnet 
simulation of two capacitively coupled LERs. The microwave transmission through the readout 
line, normalized to the maximum transmission, shows two dips that correspond to absorption by symmetric 
and antisymmetric resonant modes with frequencies separated by twice the coupling constant 
$\kappa_{i,j}$. c) Single-photon current density maps simulated with Sonnet for different 
values of the detuning between the bare LER frequencies. The modes become more localized 
as the detuning is increased.}
\label{fig:Sonnet simulations}
\end{figure*}

The platform used in the experiments, shown in Fig.~\ref{fig:set-up}a, consists of a 
superconducting chip containing a single impedance-matched transmission line and $14$ 
side-coupled Lumped Element Resonators (LERs). LERs are circuits made of an inductor $L$ and a 
capacitor $C$ connected in series, as shown in Fig.~\ref{fig:Sonnet simulations}a. The 
capacitor consists of several interdigitated fingers. Its capacitance is defined by the 
finger length, width, and separation. Regarding the inductor, a meander-like design is used. 
The inductance is mainly determined by the length and width of the superconducting metal 
stripe. 

The resonance frequency of each LER$-i$ is $\omega_{\text{r},i} =1/\sqrt{LC}$, with 
$i=1,\ldots,14$. 
Every LER in the chip has a slightly different frequency. For this purpose, all LERs have the 
same $L\simeq 6$~nH whereas $C_{i}$ is varied by changing the number of fingers and their 
length. The $7$-turn inductor is made of a $4 \mu$~m wide strip with the dimensions 
indicated in Fig.~\ref{fig:Sonnet simulations}a. The capacitors contain from $10$ to 
$26$, $6$~$\mu$m wide fingers separated by $6$~$\mu$m gaps. Their length varies from $690$ to 
$750$~$\mu$m.  The geometrical capacitance (without taking into account any parasitic 
capacitance) ranges between $0.99$~pF and $0.28$~pF. The coupling to the 
feed-line $\kappa_{\mathrm{c},i}$ depends the distance of the resonator to the transmission 
line. This chip was designed for coupling values $\kappa_{\mathrm{c},i} \approx 20 -30$~kHz.

LERs in the chip are distributed in pairs. LERs $(i,j)$ forming a pair are oriented with 
their capacitors facing each other, as illustrated by Fig.~\ref{fig:Sonnet simulations}. This 
geometry induces a capacitive coupling whose strength $\kappa_{i,j}$ is different for each 
pair and depends on the distance between them. 

\begin{table}[tb!]
%\centering
\resizebox{\columnwidth}{!}{%
\begin{tabular}{|c|c|c|c|c|c|} \hline    
LER & $\omega_{\text{r}}/2\pi$ (GHz) & $\kappa/2\pi$ (kHz) & $\kappa_{c}/2\pi$ (kHz) & Sim. $\kappa_{i,j}/2\pi$ (MHz)& Exp. $\kappa_{i,j}/2\pi$ (MHz) \\ \hline 
1 & 1.703& 91  & 18.5 & \multirow{2}{*}{\centering 2.4} & \multirow{2}{*}{\centering 1.06} \\ \cline{1-4}
2 & 1.710& 103& 7 & &\\  \hline
3 & 1.904 & 68 & 52 & \multirow{2}{*}{\centering 0.33} & \multirow{2}{*}{\centering -} \\ \cline{1-4}
4 & 1.926 & 40  & 15 & & \\ \hline
5 & 2.109 & 80  & 62 & \multirow{2}{*}{\centering 0.965}  & \multirow{2}{*}{\centering -} \\ \cline{1-4}
6 & 2.147 & 244 & 22 & &\\ \hline
7 & 2.112 & 91  & 18 & \multirow{2}{*}{\centering 30}  & \multirow{2}{*}{\centering -} \\ \cline{1-4}
8 & 2.187 & 184  & 29 &  &  \\ \hline
9 & 2.719 & 61  & 50 & \multirow{2}{*}{\centering 5.75} & \multirow{2}{*}{\centering 6.5} \\ \cline{1-4}
10& 2.729& 44  & 31 & &\\ \hline
11& 2.645 & 213  & 28 & \multirow{2}{*}{\centering 9}& \multirow{2}{*}{\centering -} \\ \cline{1-4}
12& 2.665 & 62 & 23 & & \\ \hline
13& 3.274& 185 & 22 & \multirow{2}{*}{\centering 2.66} & \multirow{2}{*}{\centering -} \\ \cline{1-4}
14& 3.291 & 130 & 29 & &\\ \hline
\end{tabular}}
\caption{\justifying Characteristic parameters of the $14$ LER resonances 
obtained by fitting microwave transmission data to the theoretical model described in 
Appendix~\ref{sec:theory}. Here, `Sim. $\kappa_{i,j}$' is the capacitive coupling strength between 
the two LERs within each pair, 
determined by simulating the microwave transmission using Sonnet, as shown in 
Fig.~\ref{fig:Sonnet simulations}b, while `Exp. $\kappa_{i,j}$' denotes the coupling strength 
obtained 
from experiments performed for LER-1 and LER-2 and for LER-9 and LER-10. For these two LER pairs, the 
other parameters have been adjusted using the experimental $\kappa_{i,j}$, whereas for the rest 
the simulated $\kappa_{i,j}$ was used.}
\label{fig:tabla_zero_field}
\end{table}

The LER and readout line geometries were designed using AutoCAD. The microwave transmission 
through the line near the LER resonances was simulated using Sonnet. When both LERs of a 
pair are identical the frequency splitting between the two resonant modes 
$\Delta \omega_{\rm r} = 2 \kappa_{i,j}$ (Fig.~\ref{fig:Sonnet simulations}b), which allows 
estimating the capacitive coupling strength. We design the circuit such that 
$\kappa_{i,j}/2 \pi$ ranges from $0.3$ to $30$~MHz. In order to allow individually addressing 
the LERs within each pair, they are detuned by approximately $20 - 30$~MHz, such that the 
`dispersive' condition $\Delta \omega_{\text{r}} > \kappa_{i,j}$ is approximately fulfilled. 
Moreover, nearest LER pairs are separated by $0.6$ mm to $1$~mm and detuned 
$250$~MHz from each other to prevent unwanted crosstalk.

The chip fabrication process starts with the pretreatment of a $275$~$\mu$~m thick Si 
substrate in a $1$~$\%$ hydrofluoric acid bath to remove the native silicon oxide layer. A 
$125$~nm thick superconducting NbTiN film is then deposited onto the Si substrate by using 
confocal DC AJA magnetron sputtering, with a chamber base pressure below 
$2\times10^{-8}$~Torr. The deposition conditions were controlled with an argon flow of 
$15$~sccm and a nitrogen flow of $1$~sccm as a plasma 
processing gases, maintaining a chamber pressure of $2$~mTorr and a power setting of $200$~W. 
These conditions lead to a deposition rate of $1.21$~$\mathring{A}/s$. Afterwards, maskless laser writer 
lithography with a MicroWriter ML3 Pro system is employed to pattern the chip 
design onto the NbTiN film. For this, a layer of negative photoresist (AZ2070) is spin-coated onto 
the surface and exposed to a $405$~nm laser 
at room temperature. Following the development of the resist, reactive ion etching is 
performed. The etching gases, SF$_6$ and Ar, were flowed at rates of $20$~sccm and $10$~sccm, respectively. 
The process operated at $75$~W of RF power and a chamber pressure of $10$~mTorr. 
The process concludes with the removal of the remaining photoresist by boiling in acetone, 
followed by rinsing with isopropanol.

\begin{figure}
\centering
\includegraphics[width=0.6\linewidth]{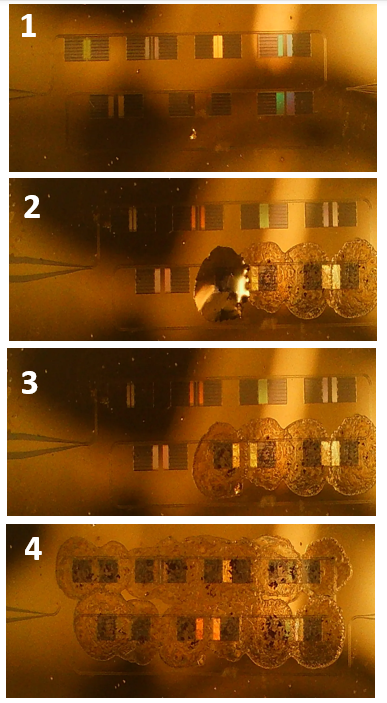}
\caption{\justifying Optical microscopy images showing different steps in the deposition of Tripak$^{-}$ 
from a $0.8$~mM solution in chlorobenzene. Starting from the bare chip (panel 1), 
in each step, a new drop is deposited on top of a LER inductor (panel 2) 
using a micropipette and then left dry under red light (panel 3) until all LER inductors are covered with the 
desired sample (panel 4). The dry deposits consist of the free radical molecules dispersed in polystyrene}
\label{fig:Tripak_deposition}
\end{figure}

\subsection{Molecular integration and characterization}\label{ssec:integration}
The PTMr and Tripak$^-$ organic free radical molecules were dissolved in chlorobenzene, 
with concentrations ranging from $0.08$ to $8$~mM, together with polystyrene. 
Then, $0.1 \mu$~l drops were transferred onto 
different parts of the circuit, namely the LER inductors or the transmission line, and left 
dry under red light, as shown in Fig.~\ref{fig:Tripak_deposition}. The resulting solid deposits 
are very stable at ambient conditions and contain a nominal spin 
number $N$ determined by concentration, namely $N \simeq 5\times 10^{12}$, 
$N \simeq 5 \times 10^{13}$ and $N \simeq 5 \times 10^{14}$ for, respectively, 
$0.08$~mM, $0.8$~mM and $8$~mM solutions. Although polystyrene prevents aggregation 
of the free radical molecules, the dry deposits are still quite inhomogeneous, 
which leads to different spin-photon couplings even for drops with nominally the same number of spins. 
In order to cover different LERs within a pair with a different free radical sample,  
the drops were landed onto the region separating two neighbor inductors from different LER pairs. 
An example is shown in Fig.~\ref{fig:autoval_LERs1and2} below.

% Aquí avanzamos el contador y registramos la etiqueta
\begin{figure}
   %% \stepcounter{figure}
\includegraphics[width=1\linewidth]{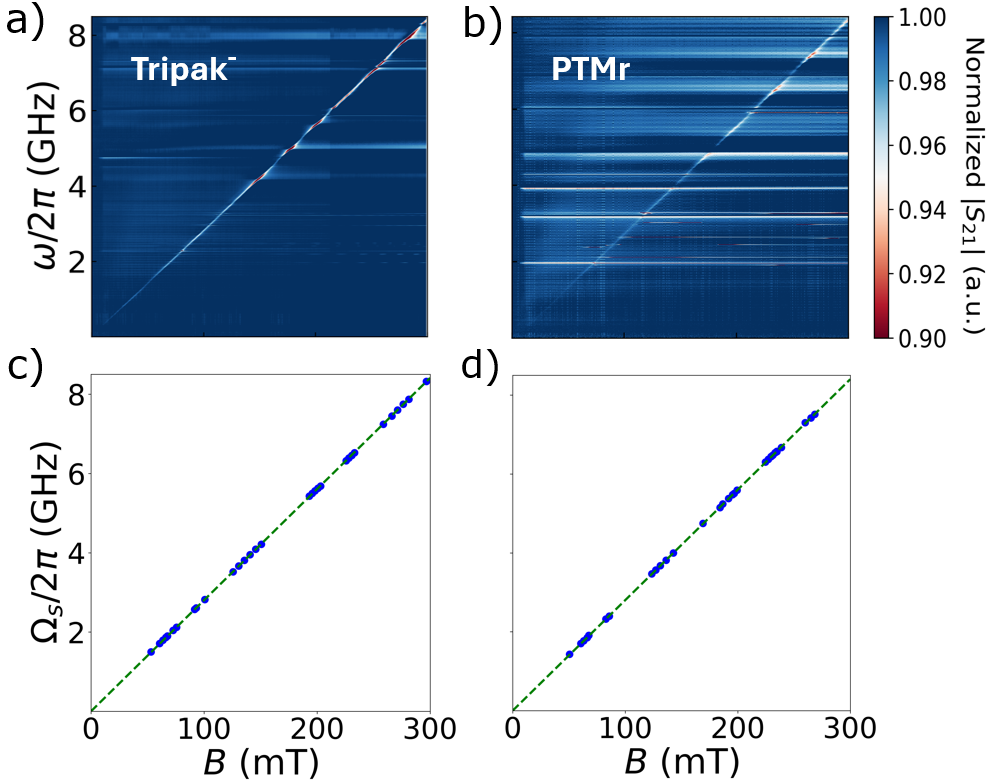}
\caption{\justifying Spin characterization by broadband magnetic spectroscopy. 
Two-dimensional plots of the microwave transmission of a superconducting waveguide coupled to 
Tripak$^{-}$ (a) and PTMr (b), measured at $T = 11$~mK as a function of frequency and 
magnetic field. Magnetic field dependence of the spin resonance frequency for Tripak$^{-}$ 
(c) and PTMr (d). The lines are least-squared fits based on $\Omega_{S}=g\mu_{B}B/\hbar$ giving $g = 2.003$ for 
Tripak$^{-}$ and $g = 2.001$ for PTMr.} 
\label{fig:Espectros}
\end{figure}

\begin{figure}
\centering
\includegraphics[width=1.0\columnwidth, angle=-0]{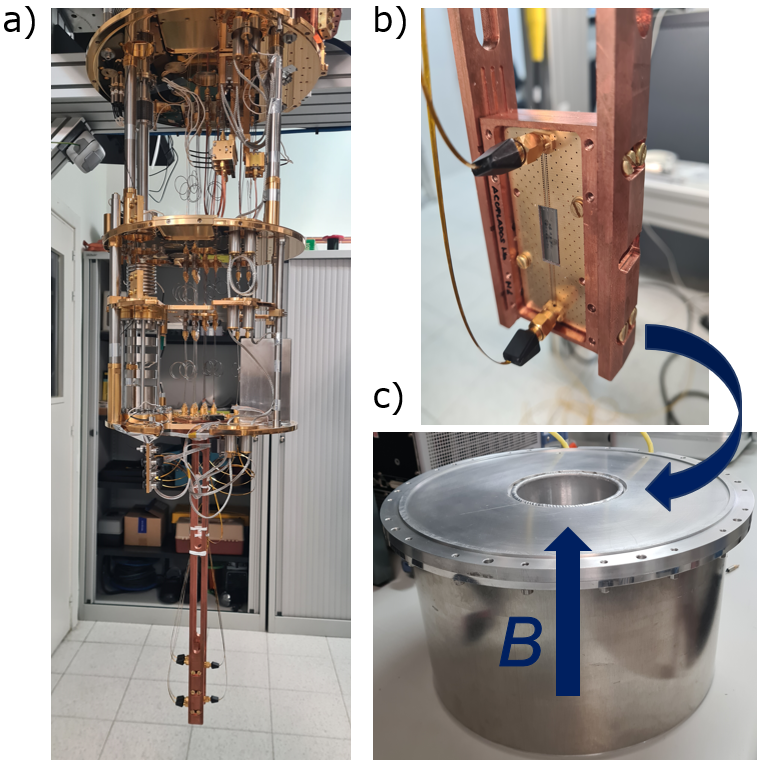}
\caption{\justifying Microwave transmission setup. a) Image of the cryo-free $^{3}$He-$^{4}$He dilution 
refrigerator used for both the cw microwave transmission experiments and the pump-probe 
experiments with microwave pulses. b) Cold finger hosting the 
superconducting chip at the center of the $1$~T cryo-free superconducting magnet c). The 
magnetic field $\vec{B}$ is applied in the chip's plane, parallel to the readout transmission line.}
\label{fig:MW_transmission_setup}
\end{figure}

The spin resonance frequencies $\Omega_{S}$ of both samples were characterized by broadband 
magnetic spectroscopy, performed by measuring the microwave transmission through a 
superconducting waveguide coupled to a molecular deposit \cite{Gimeno2023,Roca2025}. 
Results are shown in Figs. \ref{fig:spin-system}c,d and \ref{fig:Espectros}a,b. 
Figures~\ref{fig:Espectros}c and d show that $\Omega_{S} \simeq g\mu_{\rm B}SB/ \hbar$. 
From the linear slope we find $g = 2.001$ for PTMr and $g = 2.003$ for Tripak$^{-}$. 
When they are coupled to the two LERs within a pair, the slightly different magnetic responses 
of PTMr and Tripak$^{-}$ contribute to making their respective polariton frequencies tuneable 
relative to each other (Fig.~\ref{fig:spin-spin}). 

\begin{figure}[bht!]
\centering
\includegraphics[width=0.9\columnwidth, angle=-0]{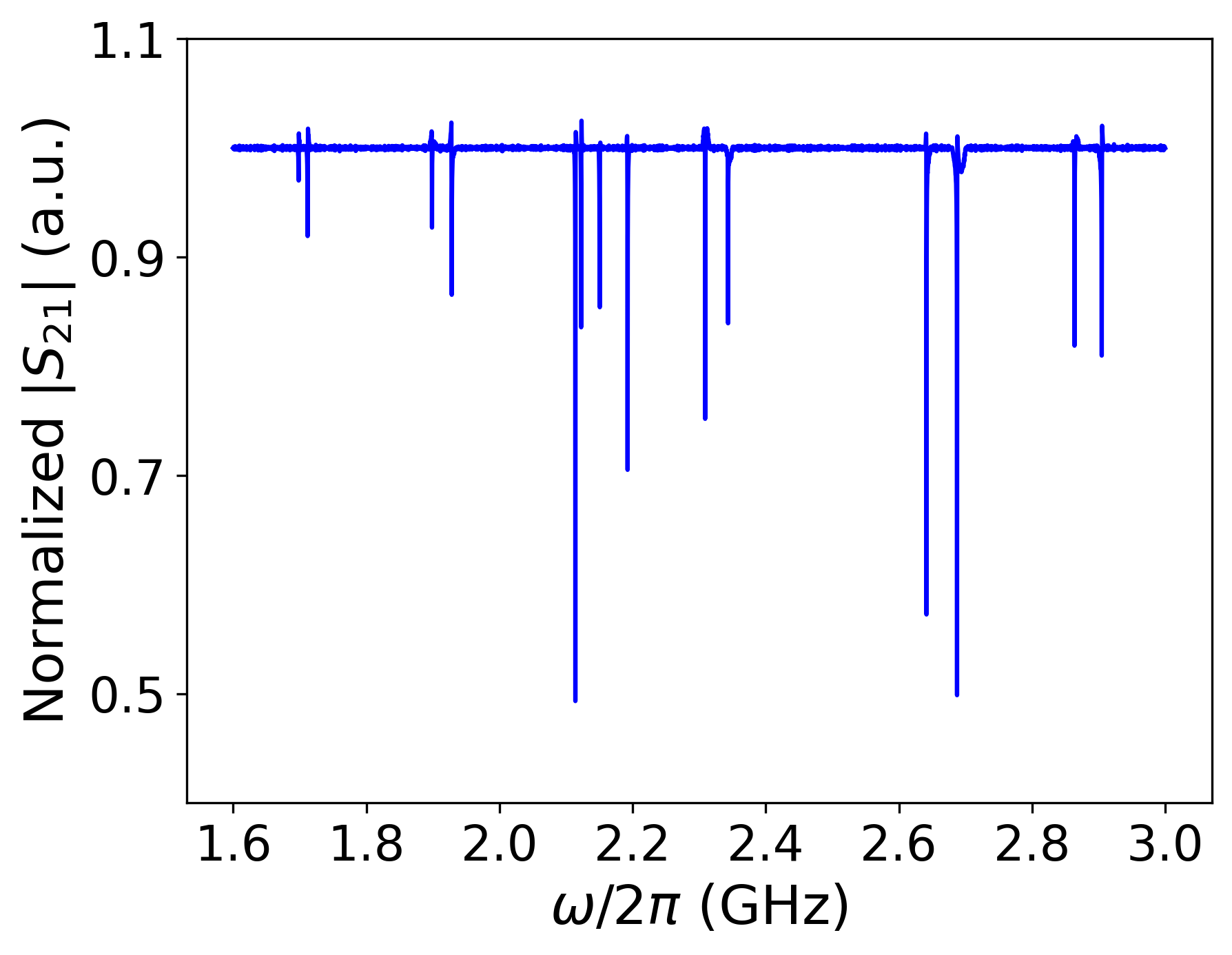}
\caption{\justifying Microwave transmission through the readout line of the device 
shown in Fig. \ref{fig:set-up}a, measured at $B=0$ and $T= 11$~mK.}
\label{fig:S21_0mT}
\end{figure}

\begin{figure*}[tb!]
\centering
\includegraphics[width=0.8\linewidth, angle=0]{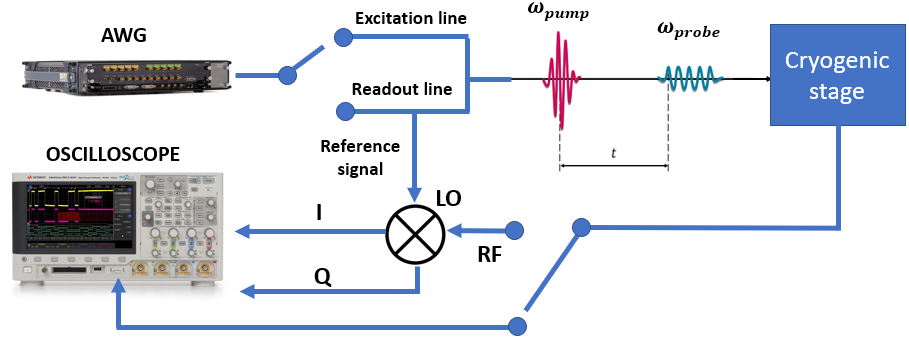}
\caption{\justifying Microwave pulse generation and detection stages. A fully digital microwave AWG 
generates the excitation and readout signals. The readout signal is first down converted by a 
synchronous $IQ$ mixer and then its two quadrature components are digitalized by a fast 
digital oscilloscope.}
\label{fig:pulsed_setup}
\end{figure*}

\subsection{Cryogenic setup}\label{ssec:cryogenic}
Experiments reported in this paper are performed at temperatures $T \geq 11$~mK, by 
thermally anchoring the chips to the mixing chamber of an $^3$He-$^4$He cryo-free dilution 
refrigerator (Fig.~\ref{fig:MW_transmission_setup}a). A copper cold finger  
(Fig.~\ref{fig:MW_transmission_setup}b) places the chip at the centre of a $1$~T axial 
superconducting magnet (Fig.~\ref{fig:MW_transmission_setup}c). The magnetic field $\vec{B}$ is applied 
along the $\hat{z}$ axis of the laboratory reference frame, which is parallel to the chip's 
readout waveguide. Cryogenic coaxial lines transmit the input and output microwave signals. 
The input signal was attenuated by two or more $-10$~dB attenuators before reaching the chip. 
The output signal was amplified $+30$~dBm at $\sim 4$~K with a low-noise cryogenic amplifier.

\begin{figure*}[bt!]
\centering
\includegraphics[width=0.9\linewidth, angle=0]{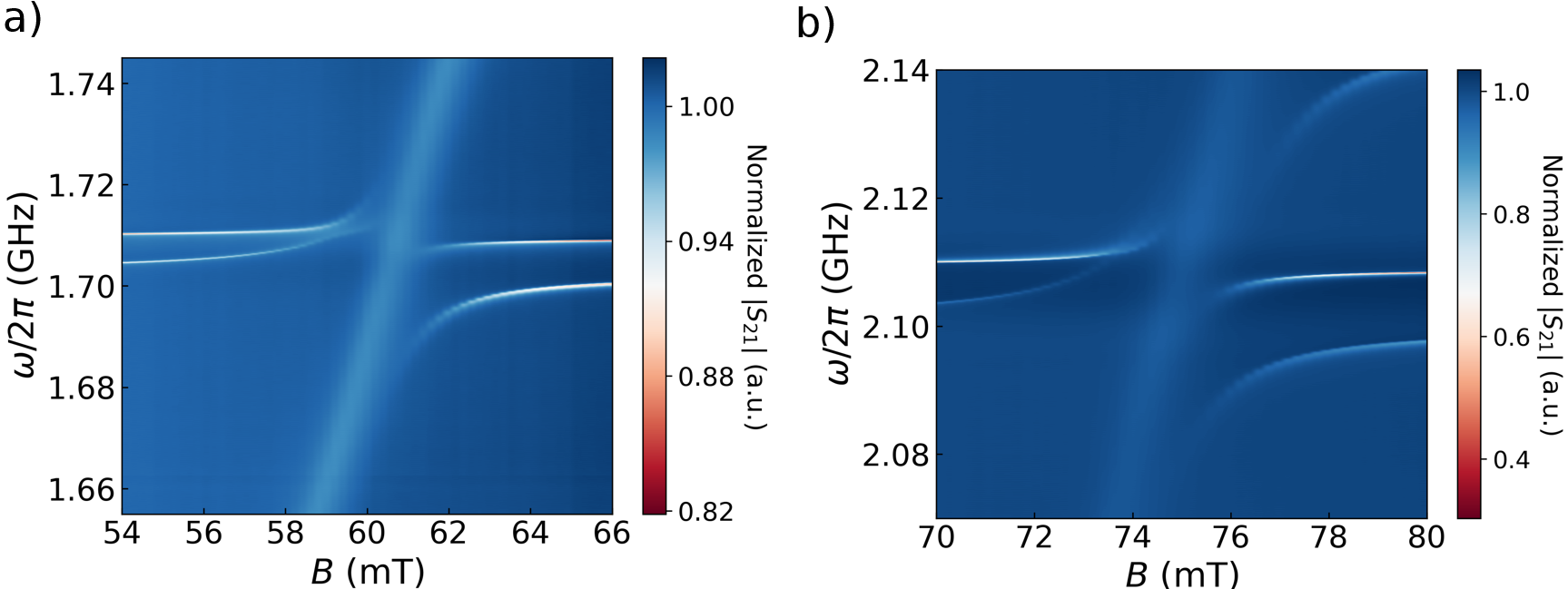}
\caption{\justifying $2D$ plot of the microwave transmission measured at $T=11$~mK near the resonance 
frequencies of: a) LER-1 (coupled to PTMr) and LER-2 (coupled to Tripak$^{-}$) belonging to a 
coupled pair; b) LER-5 (coupled to PTMr) and LER-7 (coupled to PTMr), belonging to different 
pairs, thus being mutually uncoupled. The signal arising from the direct coupling of 
spins to the readout transmission line is visible and gives in situ information on the 
spin resonance frequency.}
\label{fig:extra_CW}
\end{figure*}

\subsection{Microwave transmission experiments}\label{ssec:microwave}
The continuous wave (cw) microwave transmission through the device was 
measured using a Vector Network Analyser (VNA). The VNA generates a microwave signal, with a 
frequency $\omega/2 \pi$ ranging from $0.01$~GHz up to $14$~GHz, which is   
fed into the chip's readout line. The output signal is then amplified at $4$~K and detected 
by the VNA to determine the $S_{21}$ transmission parameter. 

Illustrative microwave transmission results, measured at $T=11$~mK and normalized to the maximum 
transmission, are shown in Fig.~\ref{fig:S21_0mT} and Table \ref{fig:tabla_zero_field}. The LER resonances give rise to 
very narrow dips in transmission that can be described by the following expression:
\begin{equation}
S_{21}=1-\frac{\kappa_{c}}{i(\omega_{\text{r}}-\omega) + \kappa}
\label{ec.1}
\end{equation}
\noindent where $\kappa$ is the LER photon decay rate and $\kappa_{c}$ is the coupling rate to 
the readout line. 

\subsection{Experiments with microwave pulses}\label{ssec:pulsed}
Figure \ref{fig:pulsed_setup} illustrates the microwave pulse generation and detection stages 
used in the pump probe experiments. It consists of a digital Arbitrary Wave Generator (AWG) 
that sends short and high power pump pulses to resonantly excite either the spins or the 
polariton modes. These are followed by a train of lower power readout pulses to detect any 
changes produced in the 
LER or polariton modes. A microwave switch, triggered by the AWG, feeds excitation and 
readout pulses into separate lines at the initial room temperature stage. The probe signal 
then splits into a local oscillator reference for the readout stage (LO) and the signal that 
is fed into the cryogenic stage. Pump and probe signals are combined again into a single 
physical line and fed through the same port of the cryostat. At the output of the cryogenic 
stage, a second switch deviates the pump pulse to a fast microwave oscilloscope in order to 
avoid damaging the detection electronics. The readout pulse is guided to an $IQ$ mixer, fed 
with the LO and RF signals, that extracts the two RF quadrature components. These components 
are then digitalized by the oscilloscope and sent to the computer for post-processing. 

\begin{figure*}[htb!]
\centering
\includegraphics[width=1.0\linewidth]{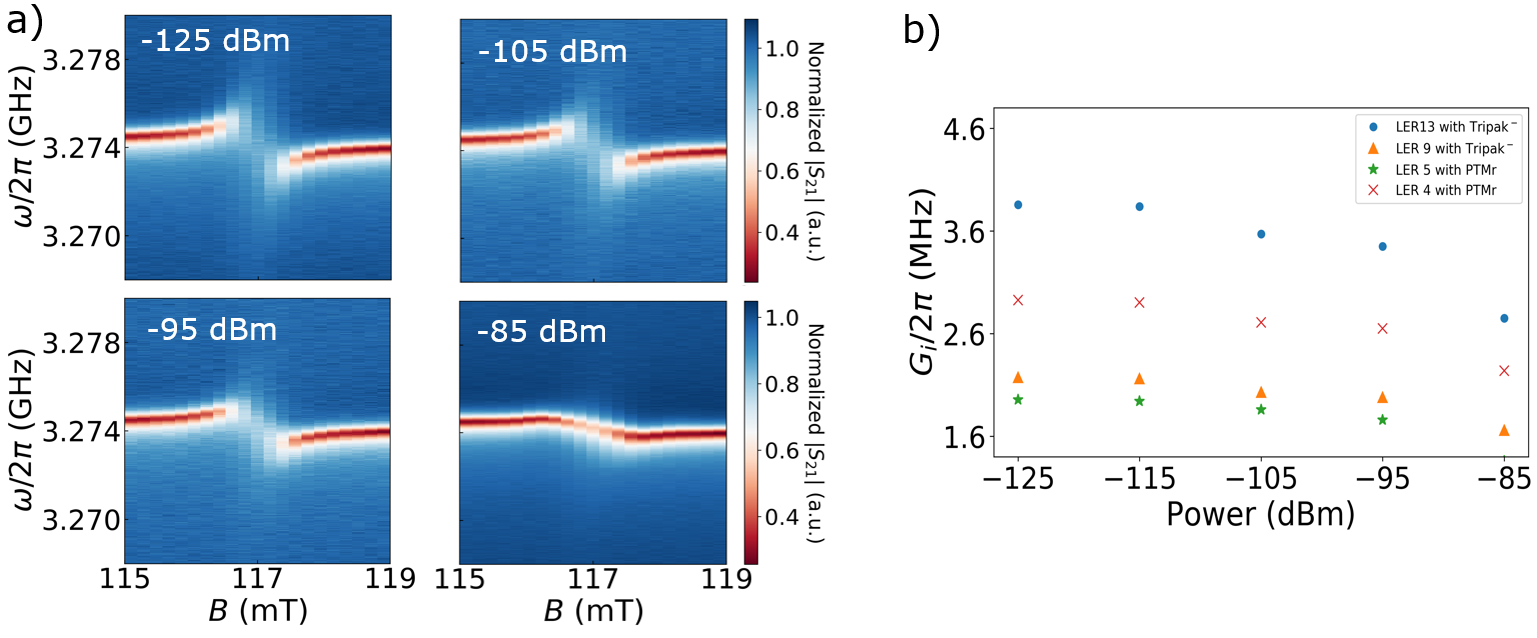}
\caption{\justifying a) $2D$ microwave transmission plots measured, at $T=11$~mK and for different input 
powers, near the resonance of LER-13 with $N\sim 5\times10^{13}$ Tripak$^{-}$ spins. The 
avoided level crossing associated with the collective spin-photon coupling gradually 
decreases as the input power is increased. b) Power dependence of the spin-photon coupling 
strength $G_{i}$ derived from microwave transmission data using input-output theory. Data are 
given for LER-13 and LER-9, hosting $N\sim 5\times10^{13}$ Tripak$^{-}$ molecules, and 
for LER-5 and LER-4, hosting $N\sim 5\times10^{13}$ PTMr molecules.}
\label{fig:LER13-powersweep}
\end{figure*}

\section{Supplementary cw microwave transmission data}\label{sec:data}
\subsection{Coupled {\em vs} uncoupled LERs}\label{ssec:coupledLER}
Figures~\ref{fig:extra_CW}a and \ref{fig:extra_CW}b show the results of cw microwave 
transmission measurements for LER-1 and LER-2, and for LER-5 and LER-7. As it is discussed in 
Section \ref{ssec:2spin} above, the former belong to a coupled pair, whereas the latter belong 
to two different 
pairs, thus they are mutually uncoupled. All these LERs host free-radical deposits with 
$N \sim 5\times10^{14}$ spins. The plots show transmission data measured over 
wider frequency and magnetic field regions than those shown in Fig. \ref{fig:spin-spin}. 
They confirm the anticrossing ocurring between the upper polariton branches of the 
two coupled LER-1 and LER-2. The gap $\Delta \omega^{+}_{\rm p}$ reflects the polariton-
polariton coupling generated by the circuit. By 
contrast, when the the polariton branches of uncoupled LER-5 and LER-7 are brought into 
resonance by the magnetic field, they simply cross, thus $\Delta \omega^{+}_{\rm p}=0$. 
These data show also that the two lower polaritons, which become visible in transmission at 
higher magnetic fields ($B \gtrsim 75$~mT), do not merge for any of these LER pairs, 
thus no polariton-polariton interaction effects on transmission are observed in this case.

\begin{figure*}[htb!]
\centering
\includegraphics[width=0.8\linewidth, angle=0]{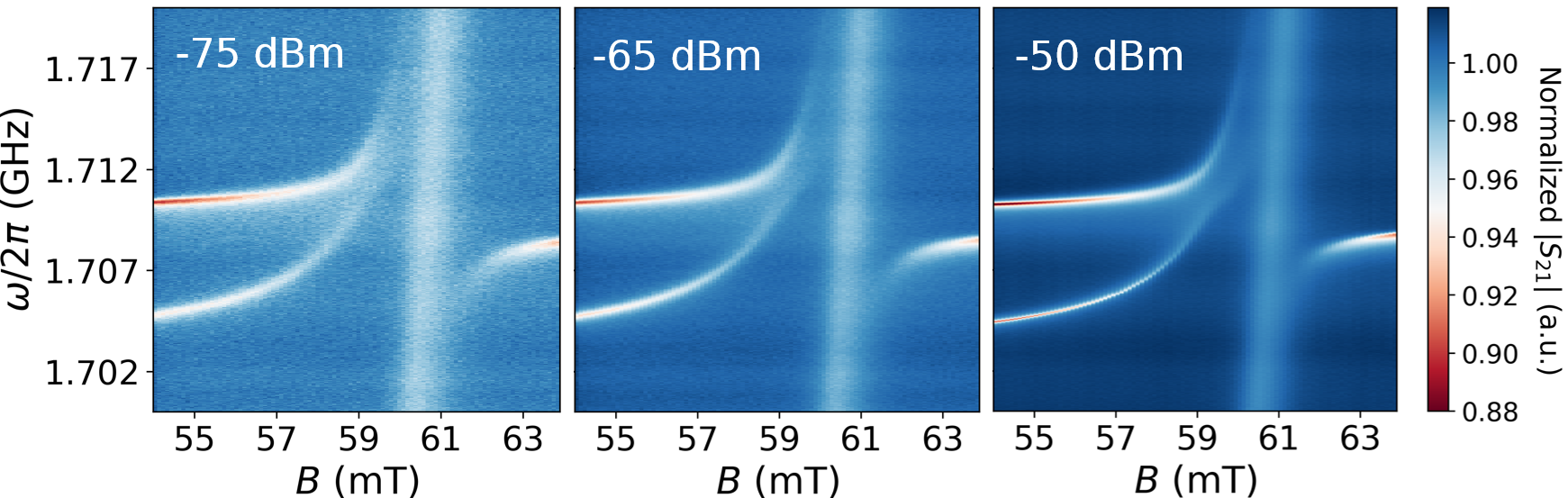}
\caption{\justifying $2D$ microwave transmission plots measured, at $T=11$~mK and for three different 
input powers, near the resonances of LER-1 and LER-2 with $N\sim 5\times10^{14}$ PTMr and 
Tripak$^{-}$ organic free-radicals, respectively. The data show also the signal arising from 
the direct coupling of the molecular spins to the readout transmission line. This signal 
bears evidence of their paramagnetic behaviour. The results show no significant changes in 
either the local spin-photon couplings or the polariton-polariton anticrossing with increasing power, 
likely because $N$ remains much larger than the number of photon excitations in both LERs.}
\label{fig:power_LERs1and2}
\end{figure*}

\subsection{Power dependence of the microwave transmission and the spin-photon coupling}\label{sec:power}
Continuous wave microwave transmission measurements were performed for different input powers. 
Illustrative results, obtained for $3.274$~GHz LER-13 coupled to 
$N\sim 5\times10^{13}$ Tripak$^{-}$ 
molecules, are shown in Fig.~\ref{fig:LER13-powersweep}a. Typical signatures characterizing  
the resonant collective coupling to the molecular spins are observed for all input powers. 
Yet, they become less visible with increasing power, suggesting that the effective 
spin-photon strength $G_{13}$ decreases. This is confirmed by the data of 
Fig.~\ref{fig:LER13-powersweep}b, which shows $G_{13}$, determined by using input-output 
theory, as a function of power. Data measured for three other LERs coupled to either 
Tripak$^{-}$ or PTMr deposits of similar size, show the same qualitative behavior. A drop in 
the effective $G_{i}$ is expected as the number of photon excitations in each LER increases, 
approaching the number of spins $N$. 

\begin{figure*}[htb!]
\centering
\includegraphics[width=\linewidth]{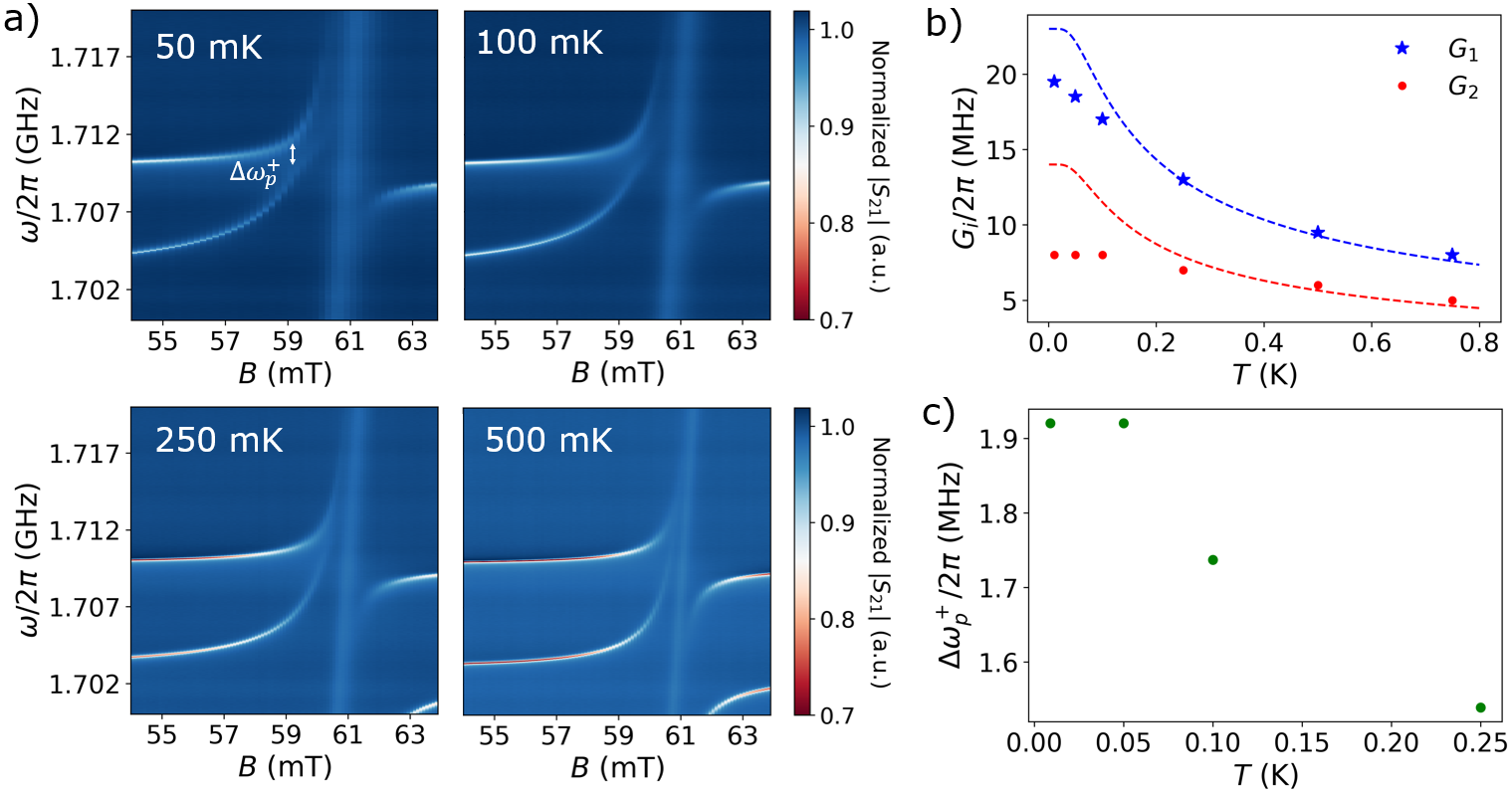}
\caption{\justifying a) $2D$ transmission plots near the LER-1 and LER-2 resonances measured at different 
temperatures. b) Temperature dependence of the collective coupling of each spin sample to 
photon modes of its host LER ($G_{1}$ and $G_{2}$ for LER-1 and LER-2, 
respectively). The lines are least-square fits based on Eq.~(\ref{eq:GvsT}), which describes 
the dependence of $G_{i}$  on spin polarization. c) Temperature dependence of the splitting 
$\Delta \omega_{\rm p}^{+}$ between the two upper polariton branches, which parametrizes the 
remote polariton-polariton coupling. Notice that, as shown in a), the avoided level 
crossing between these polariton branches shifts to higher magnetic field values as $T$ 
increases. At $T=500$~mK the anticrossing is no longer detectable.}
\label{fig:Temperature_dependence_LERs_1_and_2}
\end{figure*}

Figure~\ref{fig:power_LERs1and2} shows results measured near the resonance frequencies of 
LERs-1 and 2, which host larger ($N\sim 5 \times 10^{14}$) PTMr and Tripak$^{-}$ deposits, 
respectively. In this case, the number of photons in both LERs remains low as compared to 
$N$, and no significant effect of increasing power is observed in either $G_{1}$, $G_{2}$ or 
the polariton-polariton anticrossing $\Delta \omega^{+}_{\rm p}$.

\subsection{Temperature dependence of the microwave transmission and the spin-photon coupling}\label{sec:temperature}
The interaction between the spins and their host LERs depends on the population of the spin-
up and spin-down states, thus it is also affected by temperature. Increasing $T$ reduces the 
spin polarization at any magnetic field, leading to a decrease in $G_{i}$ given by the following expression:

\begin{equation}
G_{i}(T) = G_{i}(0) \sqrt{\tanh\left( \frac{g\mu_{\rm B}BS}{k_{\rm B}T} \right)}
\label{eq:GvsT}
\end{equation}

Figure~\ref{fig:Temperature_dependence_LERs_1_and_2}a shows microwave transmission data 
measured for the coupled LERs-1 and 2 at four different temperatures. It follows that both 
the local spin-photon coupling and the mutual polariton-polariton coupling depend on $T$. The decrease in $G_{i}$ 
with increasing $T$, shown in Fig.~\ref{fig:Temperature_dependence_LERs_1_and_2}b, also modifies the relative 
detuning between the two polariton branches, shifting the polariton-polariton anticrossing towards higher $B$. In 
addition, the gap $\Delta \omega_{\rm p^{+}}$, shown in Fig.~\ref{fig:Temperature_dependence_LERs_1_and_2}c, gets 
reduced until it becomes undetectable for $T \geq 500$~mK. The fact that the dependence of $G_{i}$ with $T$ is weaker 
than that predicted by Eq.~\eqref{eq:GvsT}, 
as shown in Fig.~\ref{fig:Temperature_dependence_LERs_1_and_2}b, suggests a non-perfect 
thermalization of the samples, in particular that of Tripak$^{-}$, below $100$~mK or the 
influence of antiferromagnetic interactions between 
the different free radicals, which is likely in concentrated samples \cite{Roca2025}.

\section{Time resolved pump-probe experiments}\label{sec:time}

\begin{figure*}[t]
\centering
\includegraphics[width=0.85\linewidth]{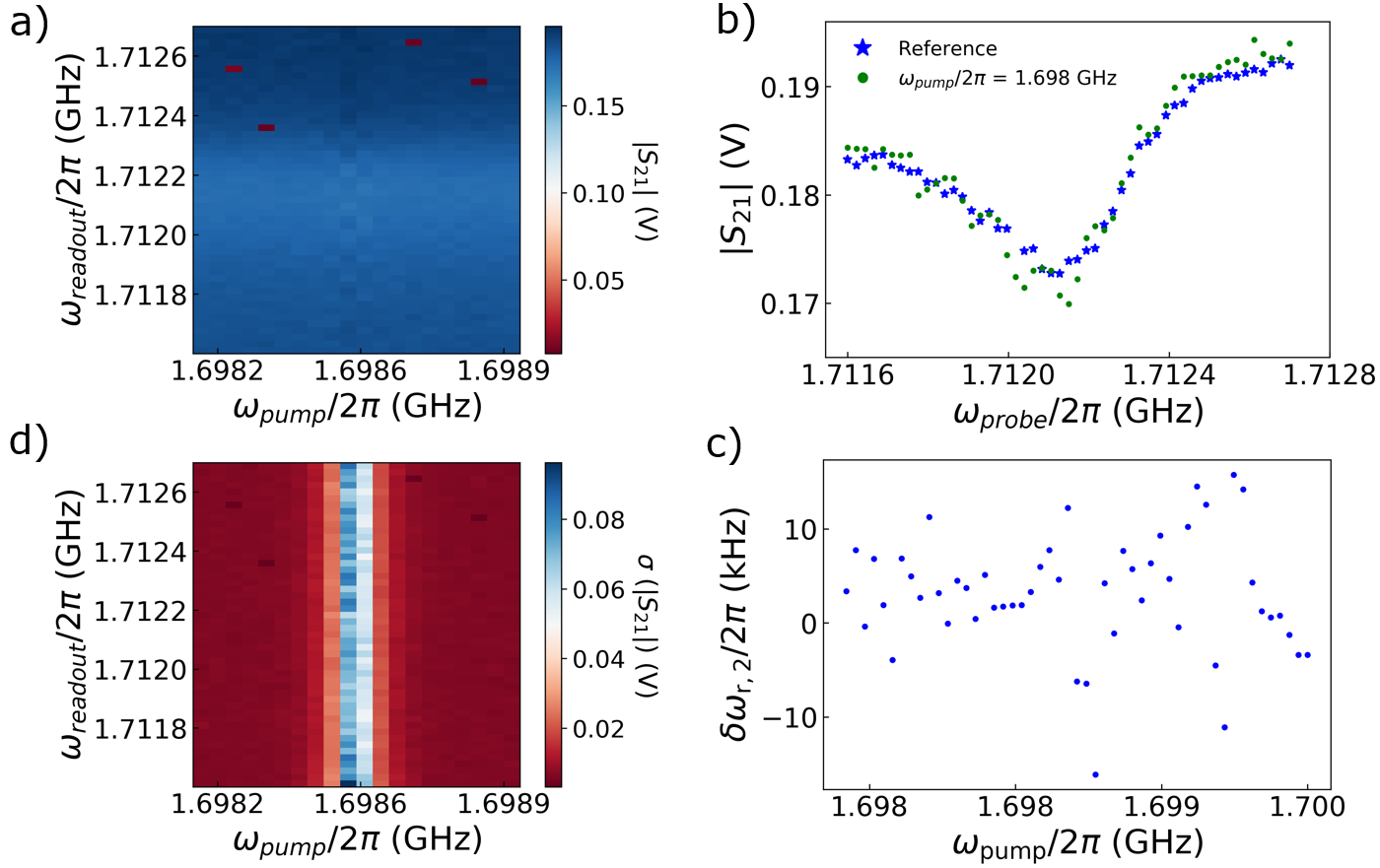}
\caption{\justifying Pump-probe photon-photon experiments performed at $B=0$ on LERs 1 and 2. a) $2D$ 
plot of the microwave transmission measured near $\omega_{\rm r,2}$ after pumping at 
frequencies close to $\omega_{\rm r,1}$. b) Example of 
LER-2 resonance measured with and without (reference) pumping on LER-1. c) Dependence of the 
LER-2 frequency shift $\delta \omega_{\rm r,2}$ as a function of the pumping frequency. d) 
Standard deviation of the transmission amplitude in a).}
\label{fig:polariton_polariton_0mT}
\end{figure*}

This section describes results of pump-probe experiments that explore three 
scenarios, corresponding to excitations of different nature: pure photons, hybrid spin-photon polaritons and 
pure spins.

\subsection{Photon-photon readout experiments}\label{ssec:photon-photon}
At $B=0$ the excitations of two mutually detuned and coupled LERs are photons localized on 
either of them. Since the two systems are purely harmonic, changes in the photon population 
of one LER shouldn't generate any shift in the resonance frequency of the other one. The 
experiments were performed on the pair formed by LER-1 and LER-2. Illustrative results 
of exciting LER-1 and reading out the frequency changes in LER-2 are shown in 
Fig.~\ref{fig:polariton_polariton_0mT}a-c. No noticeable shift in $\omega_{\rm r,2}$ is 
observed. Still, some microwave transmission oscillations appear when the pump frequency 
matches $\omega_{\rm r,1}$ (Fig.~\ref{fig:polariton_polariton_0mT}a), which become more 
noticeable in the microwave standard deviation plot shown in 
Fig.~\ref{fig:polariton_polariton_0mT}d. A plausible explanation of this effect is as 
follows. When $\omega_{\rm pump} \simeq \omega_{\rm r,1}$, LER-1 gets charged and then it 
starts decaying as soon as the pumping pulse ends. Some of the emitted photons reach LER-2 through 
the readout transmission line. However, since $\omega_{\rm r,2} \neq \omega_{\rm r,1}$, the excitation of LER-2 is 
incomplete and, besides, it oscillates with $\omega_{\rm pump} - \omega_{\rm r,2}$, which reflects itself in the 
transmission through the readout line.

\subsection{Polariton-polariton readout experiments}\label{ssec:polariton-polariton}

We next consider pump-probe experiments performed on the same pair formed by 
LER-1, coupled to $N \sim 5\times10^{14}$ PTMr molecules, and LER-2, coupled to $N\sim 5\times10^{14}$ 
Tripak$^{-}$ molecules, at $B=54$ mT and at $B=64$ mT, when the spins are still relatively 
far from resonance with either LER (see Fig.~\ref{fig:extra_CW}a). Therefore, since 
$(\omega_{\rm r,2} - \omega_{\rm r,1})/2 \pi \simeq 7$~MHz is much larger than 
$\kappa_{1,2}/2 \pi \sim 1$~MHz, the upper polaritons consist mainly of photons localized at 
either LER-1 or LER-2. However, contrary to what happens at $B=0$, the local spin-photon 
couplings $G_{1}$ and $G_{2}$ do still confer to them a tiny spin component and make their 
frequencies, $\omega_{\rm p,1}^{+}$ and $\omega_{\rm p,2}^{+}$, depend on $B$. The 
excitation of, say, the upper LER-1 polariton induces then a local spin polarization change 
that, via the capacitive coupling $\kappa_{1,2}$, must generate a shift in the frequency 
$\omega_{\rm p}^{+}$ of the upper LER-2 polariton. 

\begin{figure*}
\centering
\includegraphics[width=1.0\linewidth, angle=0]{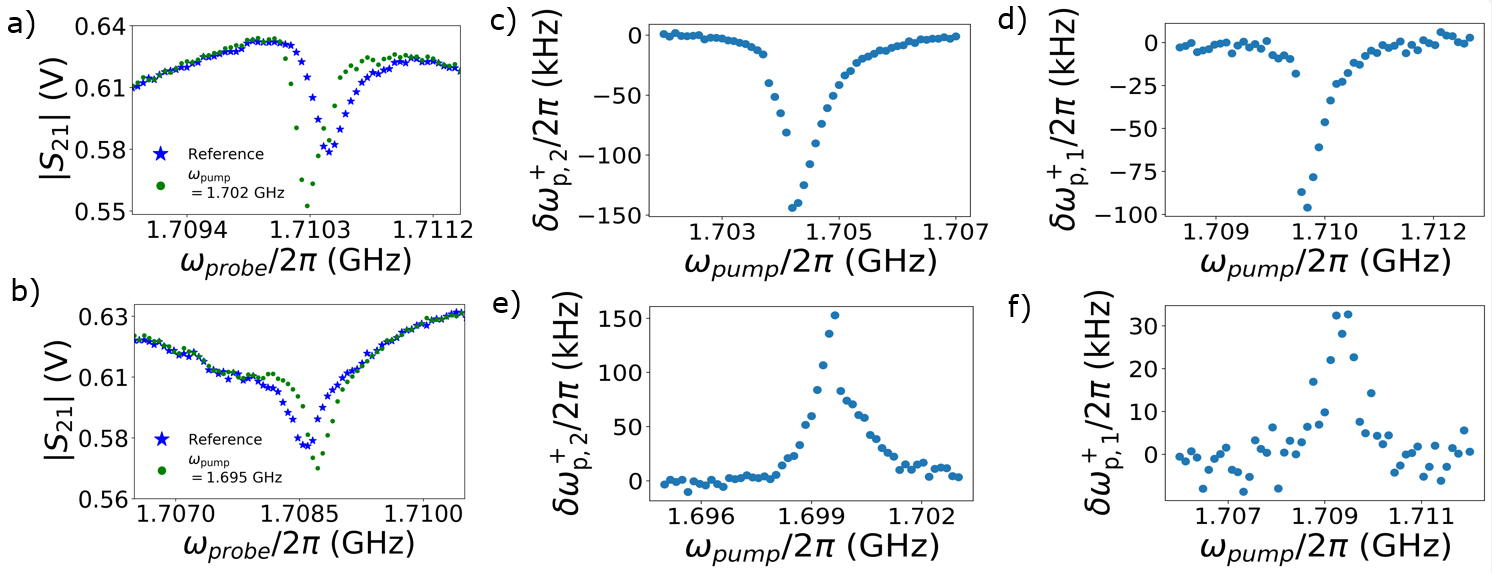}
\caption{\justifying Polariton-polariton readout measurements performed on the coupled LER-1 
(coupled to $N\sim 5\times10^{14}$ PTMr spins) and LER-2 (coupled to $N\sim5\times10^{14}$ 
Tripak$^{-}$ free radical molecules). a) and b) show the LER-2 upper polariton resonance  
measured at $B=54$ mT and $B=64$ mT, respectively, after the resonant excitation of the 
LER-1 upper polariton, compared to the reference data (measured without any previous 
excitation pulse). c) and d), spectra of the LER-1 and the LER-2 upper polaritons, 
respectively, measured at $B=54$ mT using the frequency shift 
$\delta \omega_{\mathrm{p},j}^{+}$, with $j=2,1$, respectively, of the other polariton as a probe. 
The polariton line widths derived from these data are, approximately, 
$\kappa_{\rm p, 1}/2 \pi \simeq 0.73$~MHz and $\kappa_{\rm p, 2}/2 \pi \simeq 0.34$~MHz. e) 
and f) Same as in c) and d) for $B=64$ mT. Notice the change in the sign of 
$\delta \omega_{\mathrm{p},j}^{+}$. The polariton line widths at this field are 
$\kappa_{\rm p, 1}/2 \pi \simeq 1.2$~MHz and $\kappa_{\rm p, 2}/2 \pi \simeq 0.68$~MHz.}
\label{fig:polariton-polariton}
\end{figure*}

An example of this shift is shown in Fig.~\ref{fig:polariton-polariton}a, which gives the 
LER-2 resonance recorded at $B=54$ mT before (`reference' data) and after the 
resonant excitation of LER-1 by a $50$ $\mu$s long 
$\sim -7$ dBm squared pump pulse. The readout is performed by sending a train of one hundred $20$ 
$\mu$s long pulses with frequencies spanning $\omega_{\rm p,2}^{+}$. The results show a neat 
shift in $\omega_{\rm p,2}^{+}$ with respect to the reference data. The negative sign of 
$\delta \omega_{\rm p,2}^{+}$ is due to the negative detuning of the spins with respect to LER-1 
(i.e. $\Omega_{S,1}<\omega_{\rm p,1}^{+}$, see Fig.~\ref{fig:extra_CW}a). Data measured at 
$B=64$ mT, using the same protocol at the corresponding polariton frequencies, show instead 
a positive $\delta \omega_{\rm p,2}^{+}$ (Fig.~\ref{fig:polariton-polariton}b) because then 
$\Omega_{S,1} > \omega_{\rm p,1}^{+}$. 

As discussed in Section \ref{ssec:polariton-polariton} above, the roles of the two polaritons 
can be reversed. Then, we can 
also use polaritons from LER-1 to readout polariton excitations at LER-2 and, by varying the 
excitation frequency, obtain spectroscopy information of both polaritonic branches. Results 
obtained at $B=54$ mT (leading to negative frequency shifts) are shown in Figs. 
\ref{fig:polariton-polariton}c and \ref{fig:polariton-polariton}d, whereas data  
measured at $B=64$ mT (leading to positive frequency shifts) are shown in Figs. 
\ref{fig:polariton-polariton}e and \ref{fig:polariton-polariton}f. These results confirm 
also that the polariton line width $\kappa_{\mathrm{ }p,i}$ depends on $G_{i}$ and on magnetic field, 
which combine to determine the degree of spin-photon hybridization. 

\subsection{Spin readout and relaxation measurement}\label{ssec:T1}
\begin{figure}[h!]
\centering
\includegraphics[width=0.9\linewidth, angle=0]{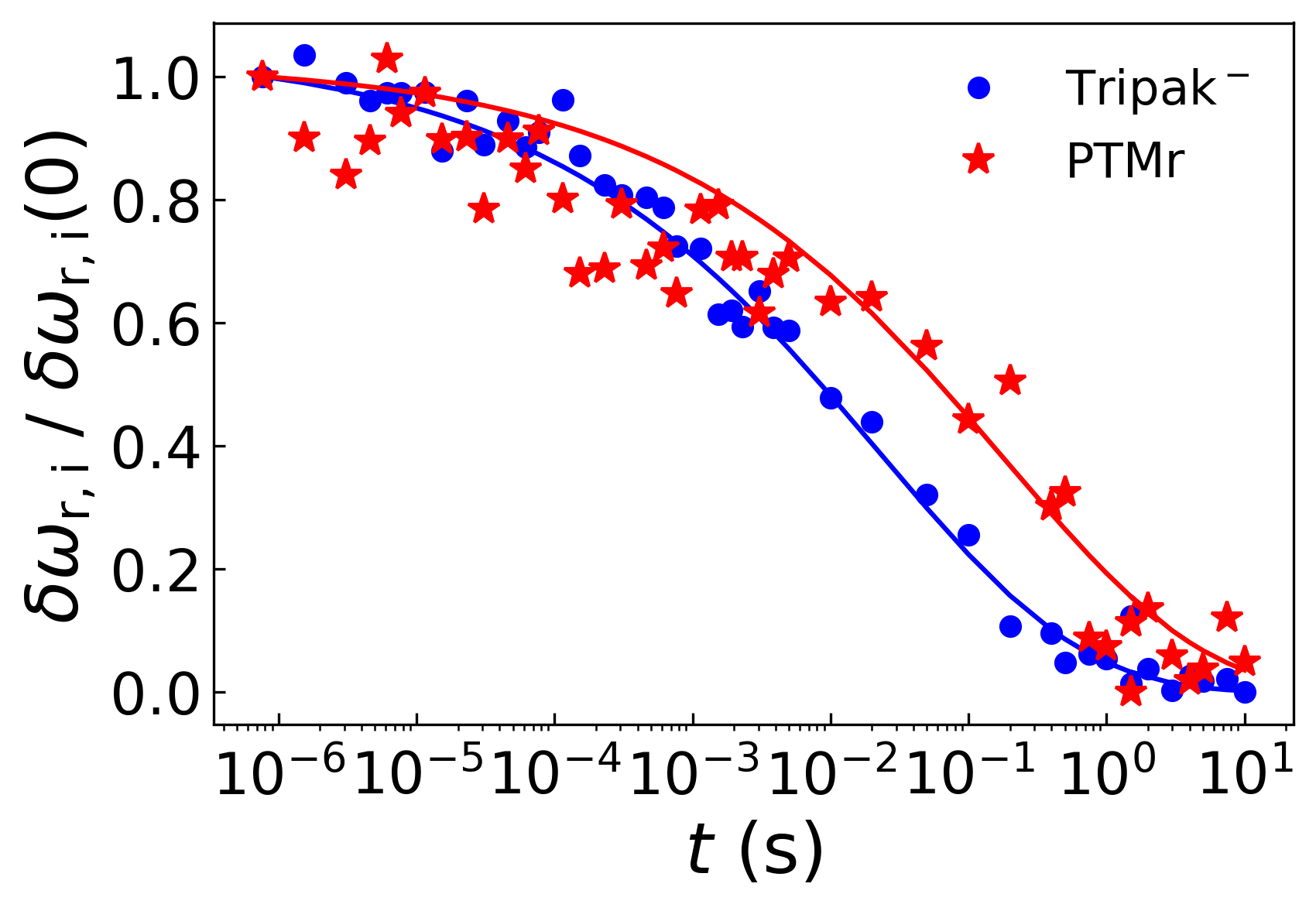}
\caption{\justifying Decay of the resonance frequency shift generated by the resonant spin excitation of 
PTMr (coupled to LER-1) and Tripak$^{-}$ (coupled to LER-2) spins at $B=66$ mT. The solid lines 
are least-square fits based on a stretched exponential [Eq. (\ref{eq:stretched})] with the parameters 
given in the text.}
\label{fig:T1}
\end{figure}

Finally, we consider the more conventional spin readout experiments, in which spin 
excitations are detected via the shift in the frequency of their host LER. Therefore, these  
experiments do not involve any non-local interaction, but are driven by the local spin 
photon couplings $G_{1}$. Experiments were performed for LER-1 and LER-2 at $B= 66$ mT, using 
the same protocol described above for the `polariton-polariton' experiments. Only, here 
the excitation frequency $\sim \Omega_{S,i}$ and the readout is performed by recording 
the resonance of the same LER-$i$. These experiments allow estimating the spin relaxation 
time $T_1$ of both samples (PTMr and Tripak$^{-}$). For this, we measure the resonance shift 
for increasing delay times $t$ between the pump and readout pulses. The results are shown in Fig.~\ref{fig:T1}.

The decay of $\delta \omega_{\rm r,i}$ with increasing $t$ is non exponential, likely 
reflecting a distribution in $T_{1}$ values associated with significant spin-spin 
interactions in these, quite concentrated, samples. In order to estimate $T_1$, we fit the 
experimental data using an stretched exponential function:

\begin{equation}
\delta \omega_{\mathrm{r},i} (t) = \delta \omega_{\mathrm{r},i} (0) e^{-(t/T_{1})^x}\ ,
\label{eq:stretched}
\end{equation}

\noindent where the exponent $x \simeq 0.3$ accounts for the distribution in relaxation times. We find  
$T_1 \simeq 0.2$ s for PTMr and  $T_1 \simeq 0.024$ s for Tripak$^{-}$.

\section{Theoretical model for polariton-polariton coupling and analysis of microwave 
transmission experiments}\label{sec:theory}

\subsection{Theoretical model: hybrid modules and their coupling}\label{sec:model}

\begin{figure*}[t]
\centering
\includegraphics[width=\linewidth]{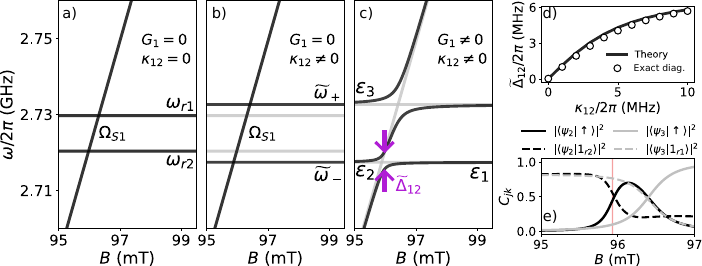}
\caption{\justifying Eigenvalues of the Hamiltonian~\eqref{eq:spin-normal-modes}
that describes two coupled LERs, one of them (LER-1, in this case with the highest 
resonance frequency $\omega_{\rm r,1}$) coupled to a spin ensemble. Panels a)–c) display 
the evolution of the system energies from the fully isolated limit to the fully interacting 
scenario. Light grey lines in b) and c) show the levels in a) and b), respectively, i.e. 
before switching a relevant interaction to better visualize its effect. 
Panel d) illustrates how the experimentally observed gap 
$\widetilde{\Delta}_{1,2}$ between the two lower polariton branches depends on the inter-
cavity coupling parameter $\kappa_{1,2}$. The open circles are calculated by exact 
diagonalization of the Hamiltonian~\eqref{eq:spin-normal-modes} and the solid line 
follows from Eqs.~\eqref{eq:g12} and \eqref{eq:predicted_gap}. Panel e) shows the probabilities 
of eigenstates $\vert \psi_2 \rangle$ (with energy $\epsilon_{2}$ in panel c) and 
$\vert \psi_3 \rangle$ (with energy $\epsilon_{3}$) in the 
single-excitation subspace basis for both LERs and the spin. The solid vertical red line 
marks the anticrossing shown by the purple arrows in panel c). The values chosen for the 
model parameters are $\omega_{\text{r}1}/2\pi=2.730$~GHz, 
$\omega_{\text{r}1}/2\pi=2.720$~GHz, $G_1/2\pi=5.4$~MHz, $\kappa_{1,2}/2\pi = 6.49$~MHz.}
\label{fig:1spin2ler_theory}
\end{figure*}

In this section, we present a theoretical framework that describes the interactions within 
a LER-spin ensemble hybrid system, its energies and wavefunctions, and how these are 
modified by its coupling to another hybrid module and by external parameters, mainly the 
magnetic field. The system under study is sketched in Fig.~\ref{fig:model}a. It 
includes a LER pair, hereafter labelled generically as LER-1 and LER-2, with bare resonance 
frequencies $\omega_{\mathrm{r},1}$ and $\omega_{\mathrm{r},2}$. Each LER can host a different spin 
ensemble, giving rise to local collective spin-photon couplings $G_{1}$ and $G_{2}$. We also denote by 
$\kappa_{\mathrm{c},1}$ and $\kappa_{\mathrm{c},2}$ 
the couplings of these LERs to the readout line and by $\kappa_{1,2}$ the 
mutual capacitive coupling between them (see Fig.~\ref{fig:Sonnet simulations}). 
The Hamiltonian describing this system is given by

\begin{equation}
\label{eq:og_full_hamiltonian}
\mathcal{H} = \mathcal{H}_0 + \mathcal{H}_{\text{sr}} + \mathcal{H}_{\text{rr}}\ ,
\end{equation}
where
\begin{equation}
\label{eq:bare_ham}
\mathcal{H}_0 = \sum_{j=1}^2
\left( \omega_{\text{r},j}a_j^\dagger a_j + \Omega_{S,j}S_j^z \right) \ ,
\end{equation}

\begin{equation}
\label{eq:spin-res-ham}
\mathcal{H}_{\text{sr}} = \sum_{j=1}^2 G_j\left(a_jS_j^+ + a_j^\dagger S_j^-\right)
\end{equation}
and
\begin{equation}
\label{eq:res-res-ham}
\mathcal{H}_{\text{rr}} = \kappa_{1,2}\left(a_1^\dagger a_2 + a_1a_2^\dagger\right)\ .
\end{equation}

In the expressions above, $a_j$ and $a_j^{\dagger}$ represent the photon annihilation and 
creation operators, respectively, in each cavity $j=1,2$, 
$S_j^\alpha$. with $\alpha = z,+,-$. represent the 
conventional spin-$1/2$ operators and the spin resonance frequency 
is related to the magnetic field through $\Omega_{S,j} = g_j\mu_\text{B}B$, in agreement 
with the experiments shown in Fig. \ref{fig:Espectros}.

Before describing situations characterized by a finite interaction between 
the two LERs, given by Eq. \eqref{eq:res-res-ham}, 
we first consider the limit of vanishing $\kappa_{1,2}$, which allows introducing mathematically 
the central single object of this work, the spin polaritons. 
These are single light and matter excitations that arise as eigenstates of the Hamiltonian formed
by $\mathcal{H}_0 + \mathcal{H}_{\text{sr}}$. In order to write these eigenstates down,
let us first define our basis as $\{|n_1,m_1^z,n_2,m_2^z\rangle\}$ where $n_j$ indicates the number of 
photons in LER $j$ (here, either $0$ or $1$) and $m_j^z = \{\uparrow, \downarrow\}$ is the spin polarization 
along its quantization axis $z$. Let $\ket{\downarrow}$ be the ground state of the spin subsystem.
Now, the polaritons of module $j\neq i$ are defined by the following states 
\begin{subequations}\label{eq:1exc_wfs}
\begin{align}
\ket{p_j^+} &= \left(\sin\frac{\theta_j}{2}\,\ket{0,\uparrow}_j + \cos\frac{\theta_j}{2}\,\ket{1,\downarrow}_j\right)\ket{0,\downarrow}_i\ , \\
\ket{p_j^-} &= \left(\cos\frac{\theta_j}{2}\,\ket{0,\uparrow}_j - \sin\frac{\theta_j}{2}\,\ket{1,\downarrow}_j\right)\ket{0,\downarrow}_i\ ,
\end{align}
\end{subequations}
\noindent and similarly for the polaritons of module $i$. Here, $\theta_j = \tan^{-1}(2G_j/\Delta_j)$ and 
$\Delta_j = (\Omega_{S,j} - \omega_{\mathrm{r},j})$ are the usual parameters that 
characterize the spin-photon hybridization within a Jaynes-Cummings 
model \cite{jaynes1963comparison}. The maximum spin-photon hybridization occurs when
$\theta_j = \pi/2$, which corresponds to the resonance condition $\Delta_j = 0$.

In the following sections, we consider systems with increasing level of complexity, which 
correspond to the two experimental situations described in Section \ref{sec:results}. The first 
includes two coupled LERs but just one spin ensemble that interacts with only one of them. 
Then, we'll discuss the more general scenario where each LER hosts a different spin 
ensemble.

%%%%%%%%%%%%%%%%%%%%%%%%%%%%%%%%%%%%%%%%%%%%%%%%%%%%%%%%%%%
%%%%%%%% FIRST EXPERIMENT: 1 SPIN AND 2 LERS %%%%%%%%%%%%
%%%%%%%%%%%%%%%%%%%%%%%%%%%%%%%%%%%%%%%%%%%%%%%%%%%%%%%%%%%

\subsubsection{First scenario: one spin ensemble and $2$ LERs}\label{sssec:1-spin-theory}

For clarity, we begin by studying the case where only one of the LERs hosts a coupled 
spin ensemble, corresponding to the setup shown in Fig.~\ref{fig:remote-spin}a. 
In this case, we set $\Omega_{S,2} = 0$ in Eq.~\eqref{eq:bare_ham} and $G_2 = 0$ in 
Eq.~\eqref{eq:spin-res-ham} and obtain the other parameters from experiment. 
Figure~\ref{fig:1spin2ler_theory} shows the results of calculations that match the most salient 
features found in the experiments. The main finding is the remote coupling $G_{1,2}$ of the 
spin ensemble to the nominally empty LER, labelled in these calculations as LER-2, which is 
mediated by the capacitive 
interaction $\kappa_{1,2}$. Our aim here is to derive an expression for $G_{1,2}$ that 
shows how this indirect coupling depends on the circuit design 
(cf. Fig.~\ref{fig:Sonnet simulations}) and then also provide information about the system 
energies and wave functions as a function of $B$. 

In order to carry out these calculations and gain some intuition on the results, we'll 
start from the unperturbed Hamiltonian $\mathcal{H}_{0}$ [Eq. \eqref{eq:bare_ham}], 
whose energy eigenstates are trivial localized photon and spin excitations 
(\ref{fig:1spin2ler_theory}a), and  
proceed by adding the different interactions. When $\kappa_{1,2} \neq 0$ 
(Figs \ref{fig:1spin2ler_theory}b and Figs \ref{fig:1spin2ler_theory}c) it is convenient to 
work in a basis defined by the normal modes $\mu = +,-$ of the coupled LER pair, described by new 
creation and annihilation operators $b_{\mu}^{\dagger}$ and $b_{\mu}$, respectively. In this new basis, the 
Hamiltonian reads

\begin{equation}\label{eq:spin-normal-modes}
\mathcal{H} = \Omega_{S,1}S_1^z + \sum_{\mu=+,-} \widetilde{\omega}_\mu b_\mu^\dagger b_\mu + \sum_{\mu=+,-} 
\widetilde{G}_{1,\mu}(b_\mu S_1^+ + b_\mu^\dagger S_1^-)\ ,
\end{equation}

\noindent with $\widetilde{\omega}_\pm = \frac{1}{2}\left(\omega_{\text{r}1}+\omega_{\text{r}2}\pm\sqrt{(\Delta\omega_{\text{r}})^2 + 4\kappa_{1,2}^2}\right)$ and 
$\Delta\omega_{\rm r} = \omega_{\text{r}2}-\omega_{\text{r}1}$. 
The normal mode operators $b_\mu$ and the single LER ones $a_{j}$ are related by
\begin{subequations}\label{eq:normal_mode_operators}
\begin{align}
b_+ &= \cos\frac{\phi}{2}\,a_1 + \sin\frac{\phi}{2}\,a_2\ , \\
b_- &= -\sin\frac{\phi}{2}\,a_1 + \cos\frac{\phi}{2}\,a_2\ . 
\end{align}
\end{subequations}
\noindent where $\phi = 
\tan^{-1}\left(2\kappa_{1,2}/\Delta\omega_{\text{r}}\right)$ is a mixing angle that 
parametrizes the degree of localization of the two normal modes. Since the eigenstates 
$\vert \psi_{j} \rangle$ of \eqref{eq:spin-normal-modes} need not be localized on either LER, from now on 
we'll label them in order of increasing energy (cf Fig. \ref{fig:1spin2ler_theory}c).

It follows from Eq.~\eqref{eq:spin-normal-modes} that the spin ensemble, even though 
localized on LER-1, now couples directly to both normal modes via $\widetilde{G}_{1,\pm}$. 
These coupling rates relate to the original $G_{1}$ as follows
\begin{subequations}\label{eq:normal_mode_couplings}
\begin{align}
\widetilde{G}_{1,+} = G_1\cos\frac{\phi}{2}\ , \\
\widetilde{G}_{1,-} = -G_1\sin\frac{\phi}{2}\ 
\end{align}
\end{subequations}
\noindent 
\noindent and give rise to avoided level crossings when the spin frequency is brought into 
resonance to any of the normal modes, as shown in Fig.~\ref{fig:1spin2ler_theory}. The 
coupling $G_{1,2}$ found experimentally (Fig.~\ref{fig:remote-spin}) is then given by 
$\widetilde{G}_{1,-}$. Using Eq. \eqref{eq:normal_mode_couplings} and the fact that 
the $b_{-}$ normal mode is still predominantly localized at LER-2, we can approximate 
\begin{equation}\label{eq:g12}
G_{1,2} = \frac{G_1}{\sqrt{2}}\sqrt{1 - \frac{\Delta \omega_{\text{r}}}{\sqrt{(\Delta\omega_{\text{r}})^2+4\kappa_{1,2}^2}}}\ ,
\end{equation}
\noindent where we have omitted the negative sign from 
Eq.~\eqref{eq:normal_mode_couplings}b since it is not relevant for the discussion that 
follows.

The remote coupling $G_{1,2}$ then arises from the local spin-photon interaction 
$G_{1}$ and the delocalization of the LER pair normal modes that is generated by 
$\kappa_{1,2}$. When $\kappa_{1,2} \gg \Delta \omega_{\mathrm{r}}$, $G_{1,2} \to G_1/\sqrt{2}$. 
This result is intuitive 
considering that, in this limit, LER excitations are fully delocalized symmetric and 
antisymmetric normal modes, thus the spin ensemble couples identically to both of them. 
As mentioned above, $G_{1,2}$ leads to the gap $\widetilde{\Delta}_{1,2}$ that opens 
between the energies $\epsilon_{1}$ and $\epsilon_{2}$ of eigenstates $\psi_1$ and $\psi_2$ 
(Fig.~\ref{fig:1spin2ler_theory}c). To a good approximation, the relation between $G_{1,2}$ 
and $\widetilde{\Delta}_{1,2}$ is simply given by 
\begin{equation}
\label{eq:predicted_gap}
\widetilde\Delta_{1,2} = \min(\epsilon_2 - \epsilon_1) \approx 2G_{1,2}\ .
\end{equation}
This relation is fulfilled over a wide range of $\kappa_{1,2}$ values, as shown by 
Fig.~\ref{fig:1spin2ler_theory}d, which compares the actual gap obtained by the exact 
diagonalization of the full Hamiltonian \eqref{eq:spin-normal-modes} to that 
derived from Eqs.~\eqref{eq:g12} and \eqref{eq:predicted_gap}. From a practical point of view, 
Eq.~\eqref{eq:predicted_gap} provides a simple method to estimate the remote coupling 
constant $G_{1,2}$ and, from it, $\kappa_{1,2}$ via Eq.~\eqref{eq:g12}, using microwave 
transmission data as it is done with the data of Fig.~\ref{fig:remote-spin}.

\begin{figure}[t!]
\centering
\includegraphics[width=0.95\columnwidth]{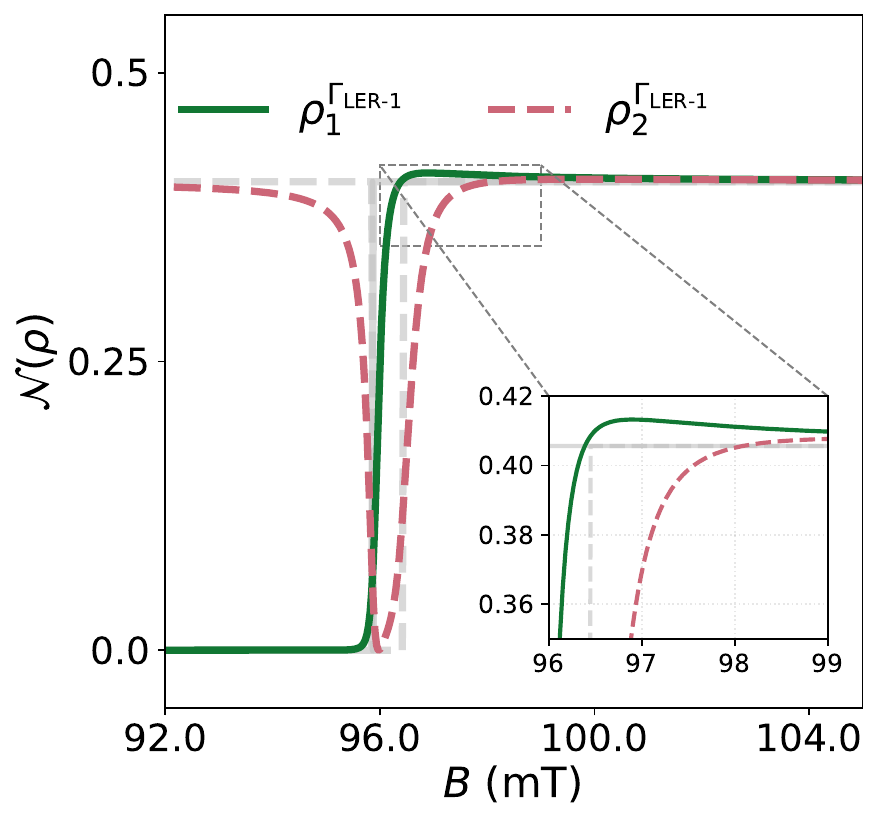}
\caption{\justifying Entanglement between LER-1 and LER-2 photons as a function of magnetic field. It 
is computed as the negativity of LER-1, defined by Eq.~\eqref{eq:def_negativity}, 
in eigenstates $\vert \psi_{1} \rangle$ and $\vert \psi_{2} \rangle$ of the 
Hamiltonian (\eqref{eq:spin-normal-modes}. The 
parameters considered are those used in Fig.~\ref{fig:1spin2ler_theory}. The light 
gray lines show the negativity calculated for $G_1=0$, i.e. when the photons are decoupled 
from the spins. The inset shows an zoom of the region close to the anticrossing between 
$\psi_1$ and $\psi_2$ in Fig.~\ref{fig:1spin2ler_theory}. A maximum entanglement between 
both photons then occurs for $G_{1} \neq 0$ near the anticrossing between the LER-1 lower 
polariton and the quasi LER-2 mode $b_{-}$ (see Fig.~\ref{fig:1spin2ler_theory}c).}
\label{fig:1s2r_photon_entanglement}
\end{figure}

It is also of interest to estimate the magnetic field at which the anticrossing occurs. 
In essence, it arises from the resonance, induced by $B$, between the 
lower polariton generated at LER-1 and the $b_{-}$ mode that is closer to LER-2. Based on 
Hamiltonian~\eqref{eq:spin-normal-modes}, we can restrict our analysis to the single-excitation 
subspace and treat the coupling $\widetilde{G}_{1,-}$ as a perturbation. Consequently, we 
re-diagonalize the block corresponding to the symmetric photonic mode $b_{+}$ and the spin 
ensemble, mutually coupled via $\widetilde{G}_{1,+}$ at LER-1, in order to extract its two 
polariton branches. The resulting polariton `pseudo-eigenenergies' are then approximately 
given by
\begin{equation}
\label{eq:pseudo-eigen-1s2l}
\tilde\epsilon_{\pm} = \frac{1}{2}\left(\Omega_{S,1} + \widetilde{\omega}_+ \pm \sqrt{(\Omega_{S,1} - \widetilde{\omega}_+)^2 + 4\widetilde{G}_{1,+}^2}\right)
\end{equation}

\noindent Since $\widetilde{\omega}_- < \widetilde{\omega}_+$, the LER-1 polariton branch 
that intersects with $\widetilde{\omega}_-$ is $\tilde\epsilon_-$. The intersection occurs 
when  $\tilde\epsilon (\Omega_{S,1}) = \widetilde{\omega}_-$ that leads to
\begin{equation}
\label{eq:crossing_field_1s2l}
\Omega_{S,1} = g_{1}\mu_\text{B}B \approx \widetilde{\omega}_- + \frac{\widetilde{G}_{1,+}^2}{\widetilde{\omega}_+ - \widetilde{\omega}_-}\ ,
\end{equation}
provided that $(\Omega_{S,1} - \widetilde{\omega}_+)\gg \widetilde{G}_{1,+}$. The  spin-photon hybridization at LER-1 
then shifts the resonance field from that of the bare spin, defined by $\Omega_{S,1} = \widetilde{\omega}_{-}$. 
Besides, this shift decreases as either the detuning $\Delta \omega_{\rm r}$ between 
the two LERs or their mutual coupling $\kappa_{1,2}$ increase. This can be understood as a 
result of the LER-1 lower polariton branch becoming progressively closer to a pure spin 
excitation at its crossing with $b_{-}$, as the two modes become further apart in energy.

\begin{figure*}[tb!]
\centering
\includegraphics[width=\linewidth]{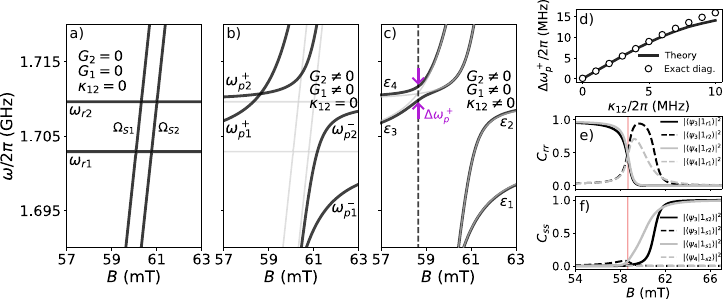}
\caption{\justifying Eigenvalues of the Hamiltonian~\eqref{eq:og_full_hamiltonian}
that describes two coupled LERs, each of them coupled to a different spin ensemble. 
Panels a)–c) display the evolution of the system energies from the fully isolated limit 
to the fully interacting scenario (corresponding to the experimental configuration 
shown in Fig.~\ref{fig:spin-spin}). Panel d) 
illustrates the relationship between the experimentally observed gap 
$\Delta \omega_{\rm p}^{+}$ between 
the two upper polariton branches and the inter-cavity coupling parameter $\kappa_{1,2}$.
The circles are obtained by exact diagonalization of the full Hamiltonian 
\eqref{eq:og_full_hamiltonian} and the solid line follows from Eqs.~\eqref{eq:polariton_gap}. 
Finally, panels e) and f) show the probabilities of eigenstates 
$\vert \psi_3 \rangle$ and $\vert \psi_4 \rangle$ within the 
single-excitation subspace basis for the two LERs (panel e) and the two spin ensembles (panel f). 
The vertical red line indicates the anticrossing point marked by a black dashed line in 
panel c). The model parameters are $\omega_{\text{r}1}$=1.7029 
GHz, $\omega_{\text{r2}}$=1.7096~GHz, $G_1 = 19.5$~MHz, $G_2=8.5$~MHz and 
$\kappa_{1,2} = 1.06$~MHz.}
\label{fig:2spin2ler_theory}
\end{figure*}

As a somewhat related effect, we next consider the entanglement between 
LER-1 and LER-2 photons in each eigenstate of the system $\vert \psi_{j} \rangle$ 
and how it evolves as a function of magnetic field. Here, we quantify this 
entanglement by using the negativity of LER-1, which is an entanglement monotone \cite{vidal2022computable} and is defined as follows 
\begin{equation}
\label{eq:def_negativity}
\mathcal{N}(\rho_j) = \frac{||\rho_j^{\Gamma_A}||_1-1}{2} = \left|\sum_{\lambda_i<0}\lambda_i\right|\ ,
\end{equation}
where $\rho_j^{\Gamma_A} = \rho_j^{\Gamma_{\rm LER-1}}$ is the partial transpose of the reduced density matrix 
$\rho_j$ with respect to subsystem LER-1. Concurrently, such reduced density matrix
is obtained by tracing out the spin (S1) degrees of freedom with respect to the subsystem 
defined by the two cavities, $\rho_j = \text{Tr}_\text{S1}(|\psi_j\rangle\langle\psi_j|)$.
The $\lambda_i$ elements in the right-most expression
refer to the eigenvalues of $\rho_j^{\Gamma_A}$.

The results, shown in Fig.~\ref{fig:1s2r_photon_entanglement}, reveal that the hybrid 
character of the lower LER-1 polariton, generated by the local spin-photon coupling 
$G_{1}$, leads to a local entanglement maximum near the anticrossing of this polariton with 
mode $b_{-}$, which is close to LER-2. As discussed above, the hybridization introduces 
some photonic character into the polariton branch, which then enhances the photon 
entanglement with respect to that given solely by $\kappa_{1,2}$ (grey lines in Fig.~\ref{fig:1s2r_photon_entanglement}). The relative intensity of this effect diminishes as 
$\kappa_{1,2}$ increases, and can thus be tuned by circuit design. The higher the baseline 
photon entanglement, the smaller the spin contribution, provided the system is not in the 
regime where $G_1 \gg \kappa_{1,2}$. The experimental example considered here represents a 
situation dominated by a relatively large $\kappa_{1,2} \simeq 0.7 \Delta \omega_{\rm r}$. 
The next section covers a different situation in a different parameter regime. 

%%%%%%%%%%%%%%%%%%%%%%%%%%%%%%%%%%%%%%%%%%%%%%%%%%%%%%%%%%%
%%%%%%%% SECOND EXPERIMENT: 2 SPINS AND 2 LERS %%%%%%%%%%%%
%%%%%%%%%%%%%%%%%%%%%%%%%%%%%%%%%%%%%%%%%%%%%%%%%%%%%%%%%%%

\subsubsection{Second scenario: two spin ensembles and 2 LERs}\label{sssec:2-spin-theory}

We next analyze the behavior of a system where both LERs host spin ensembles.
The two ensembles possess different resonance frequencies $\Omega_{S,j}$ 
due to their different $g$-factors. Moreover, each sample presents also a different spin-photon 
coupling $G_{j}$ to its host LER. The parameters considered 
here correspond to those extracted from the experiment described in 
Fig. \ref{fig:spin-spin}. The main experimental phenomenon we aim to 
characterize is the anticrossing occurring between the two upper polariton branches.

Like in the previous case, the eigenstates and the frequencies of the two coupled 
hybrid spin-cavity modules 
have been calculated by diagonalization of Hamiltonian \eqref{eq:og_full_hamiltonian}. 
The results are shown in Figure~\ref{fig:2spin2ler_theory}, which again illustrates how 
they evolve as the different interactions are `switched on'. The anticrossing, 
characterized by the $\Delta \omega_{\rm p}^{+}$ gap in Fig.~\ref{fig:2spin2ler_theory}c, 
reflects the interaction between two hybrid spin-photon excitations, 
each arising from one of the two spin-LER subsystems. This constitutes a generalization of 
the previous case, where now the capacitive photon-photon coupling $\kappa_{1,2}$ induces 
interactions between distant spins, alongside interactions between the spin system $1$ and 
LER-2, and viceversa. 

To understand the relationship between the gap and $\kappa_{1,2}$, we revisit 
the model defined by Eq.~\eqref{eq:og_full_hamiltonian}, but now with $\Omega_{S,2}$ and 
$G_2 \neq 0$. A further distinction from the previous case is that we now proceed by first 
diagonalizing the blocks corresponding to the spin-photon interaction within each 
LER, giving rise to polariton states $\vert p_j^\pm \rangle$, $j=1,2$, defined by
Eqs.~(\ref{eq:1exc_wfs}a-b) with frequencies 
$\omega_{\rm p}^{\pm}$ shown in Fig.~\ref{fig:2spin2ler_theory}b. Then, we treat the 
inter-LER coupling as a perturbation. From these, we can define
the effective coupling parameter between the two upper, symmetric, polaritons 
as the matrix element of the perturbation $\mathcal{H}_\text{sr}$ between the two states,
i.e.  $J^{+}_{1,2} \equiv \langle p_1^+|\mathcal{H}_\text{sr}|p_2^+\rangle$.

The experimentally observed gap $\Delta \omega_{\rm p}^{+}$ between the frequencies 
of these polariton branches can then be approximated by
\begin{equation}\label{eq:polariton_gap}
\Delta \omega_{\rm p}^{+} \simeq 2J^{+}_{1,2} 
= 2\kappa_{1,2}\cos\frac{\theta_1}{2}\cos\frac{\theta_2}{2} \ .
\end{equation}
\noindent The gap scales as $\kappa_{1,2}$, reflecting the dominant photonic nature of the 
polaritons, but depends also on their spin components, defined by $\theta_{1}$ 
and $\theta_{2}$, which are induced by the local spin-photon interactions $G_{1}$ and 
$G_{2}$. In fact, it is their spin component that renders them sensitive to the magnetic 
field and ultimately drives the crossing. Note that $J^{+}_{1,2}$ is not purely proportional 
to $\kappa_{1,2}$, as can be seen in Fig.~\ref{fig:2spin2ler_theory}d, since $\kappa_{1,2}$ 
also determines the magnetic field at which the two polariton branches cross. In this case, 
the approximation breaks down earlier than for Eq.~\eqref{eq:predicted_gap} due to the 
presence of additional interacting subsystems with non-negligible couplings.

The combination of local and remote interactions leads to correlations between the 
spin and photon components, which are tuned by the magnetic field as it brings 
local polaritons in and out mutual resonance. This effect can be seen by looking 
at the probabilities that the exact eigenstates $\vert \psi_3 \rangle$ and 
$\vert \psi_4 \rangle$ (again sorted by energy) 
have in the basis of local single photon and single spin excitations. These 
probabilities are shown in panels 
e) and f) of Fig.~\ref{fig:2spin2ler_theory}, respectively. The 
vertical red line corresponds to the anticrossing marked as a dashed black line in 
panel c). At this point, the two wavefunctions become maximally delocalized over the two 
LER photons. The different spin-photon couplings at each LER lead to asymmetric 
correlations between their photon excitations. For instance, panel e) shows that the mode 
originating as LER-1 acquires significantly more LER-2 character than the converse. 
Interestingly, the mode originating as LER-2 exhibits less hybridization with LER-1 
because it simultaneously transitions into a `spin-1' excitation. Here, the level crossings 
do not merely involve a predominantly photon-like eigenstate transitioning into a spin-like 
excitation, as in a conventional polariton. Rather, an eigenstate starting as a LER-$j$ 
photon converts after the crossing into an excitation of the spin ensemble $i$ located at 
the other LER-$i$. At the intermediate point, the wavefunctions exhibit non-zero amplitudes 
for all four subsystems.

To better understand this behavior and quantify the effective spin-spin interaction 
mediated by the circuit, we move to a different basis to that used in 
Eqs.~\eqref{eq:1exc_wfs}a–d. Specifically, we return to the photonic normal mode basis [cf. Eq.~\eqref{eq:spin-normal-modes}] and treat the spin-mode coupling as a perturbation. Then, 
we now define
\begin{equation}
\label{eq:h0_normalmodes}
\mathcal{H}'_0 = \sum_j \Omega_{Sj}S_j^z + \sum_{\mu=+,-}\widetilde{\omega}_\mu b_\mu^\dagger b_\mu
\end{equation}
\noindent and
\begin{equation}
\label{eq:spin-normal-mode-term}
\widetilde{V} = \sum_{j,\mu}\widetilde{G}_{j,\mu}(b_\mu S_j^+ + b_\mu^\dagger S_j^-)\ ,
\end{equation}
\noindent where the normal mode operators $b_\pm$ have been defined in 
Eqs.~\eqref{eq:normal_mode_operators}a-b, their couplings to spin-1 
$\widetilde{G}_{1,\pm}$ are those defined in Eqs.~\eqref{eq:normal_mode_couplings}a-b and 
their couplings to spin-2 read
\begin{subequations}\label{eq:spin2-normalmodes-coupling}
\begin{align}
\widetilde{G}_{2,+} = G_2\sin\frac{\phi}{2}\ , \\
\widetilde{G}_{2,-} = G_2\cos\frac{\phi}{2}\ .
\end{align}
\end{subequations}

From here, it is possible to derive an effective Hamiltonian by applying a Schrieffer-Wolff 
transformation, defined by $\mathcal{H}_{\text{eff}} = e^{\mathcal{S}}\mathcal{H}e^{-\mathcal{S}}$ 
\cite{schrieffer1966relation}. The operator $\mathcal{S}$ is such that the relation 
$\widetilde{V} + [\mathcal{S}, \mathcal{H}'_0] = 0$ is fulfilled. After this transformation, 
the effective Hamiltonian reads

\begin{equation}\label{eq:schrieffer-wolff}
\mathcal{H}_{\text{eff}} = \mathcal{H}'_0 + 
\frac{1}{2}[\mathcal{S},\widetilde{V{}}] + \frac{1}{3}[\mathcal{S}, [\mathcal{S}, \widetilde{V}]] + \dots
\end{equation}

\begin{figure}[t!]
\centering
\includegraphics[width=0.95\columnwidth]{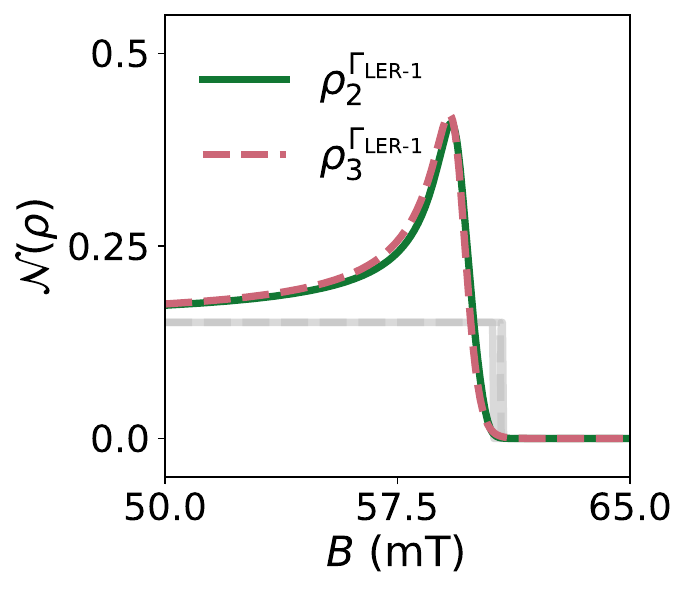}
\caption{\justifying Photon entanglement as a function of the external magnetic field computed as the negativity for the 
subsystem LER-1 according to Eq.~\eqref{eq:def_negativity}, after tracing out the spin degrees of freedom for subsystems 
$\text{s1}$ and $\text{s2}$. The parameters considered are the same as in Fig.~\ref{fig:2spin2ler_theory}. The gray 
lines indicate the negativity evolution for the same eigenstates as the respective colored lines but in the scenario 
where $G_1 = G_2 = 0$.}
\label{fig:2s2r_photon_entanglement}
\end{figure}

\noindent For the present case, the anti-hermitian operator $\mathcal{S}$ reads
\begin{equation}
\mathcal{S} = \sum_{j,\mu}\frac{\widetilde{G}_{j,\mu}}{\widetilde{\Delta}_{j,\mu}}(b_\mu 
S_j^+ - b_\mu^\dagger S_j^-)\ ,
\end{equation}
with $\widetilde{\Delta}_{j,\mu} = \Omega_{S,j} - \widetilde{\omega}_\mu$. In the resulting 
effective Hamiltonian, we are going to eliminate, up to first order in the 
series \eqref{eq:schrieffer-wolff}, the interaction between the spins and the normal modes. 
This approximation will properly describe the system whenever 
$G_j\ll\widetilde{\Delta}_{j,\mu}$, i.e. in the dispersive regime. If we stop the 
series \eqref{eq:schrieffer-wolff} at the second order, the effective Hamiltonian takes the 
form
\begin{equation}
\label{eq:eff_hamiltonian}
\mathcal{H}_{\text{eff}} = \mathcal{H}'_0 + \mathcal{H}_{\text{disp}} + \mathcal{H}_{\text{spin-spin}}\ .
\end{equation}
The first term, $\mathcal{H}'_0$, is just Eq.\eqref{eq:h0_normalmodes}, the second one,
\begin{equation}
\label{eq:disp_term}
\mathcal{H}_{\text{disp}} = \sum_{j,\mu}\frac{\widetilde{G}_{j,\mu}^2}{\widetilde{\Delta}_{j,\mu}}(2b_\mu^\dagger b_\mu + 1)S_j^z\ ,
\end{equation}
describes the usual dispersive effect of the spin-normal mode couplings on the 
frequencies of both, and the third and last term,
\begin{equation}
\label{eq:spinspin_term}
\mathcal{H}_{\text{spin-spin}} = \frac{J}{2}(S_1^+S_2^- + S_1^-S_2^+)\ ,
\end{equation}
\noindent where $J = \sum_\mu  \widetilde{G}_{1,\mu}\widetilde{G}_{2,\mu}\left(\widetilde{\Delta}_{1\mu}^{-1}+\widetilde{\Delta}_{2\mu}^{-1}\right)$, 
describes an emerging spin-spin interaction. Note that $J\propto G_1G_2\kappa_{1,2}$ since 
it is a third-order process. The spin ensemble $j$ first couples to its host LER-$j$, via 
$G_j$, which in turn forms a hybrid mode with the adjacent resonator through $\kappa_{1,2}$, 
and then interacts with the distant spin ensemble, $i$, via $G_i$. It is important to 
emphasize that this analysis holds only in the dispersive regime where 
$G_j\ll\widetilde{\Delta}_{j,\mu}$. To study the spin and photon hybridization near 
resonance, it is necessary to resort to the numerical diagonalization methods 
illustrated in Fig.~\ref{fig:2spin2ler_theory}.

Finally, we revisit the evolution of photon entanglement between the two cavities as a 
function of magnetic field, now in the presence of two spin ensembles. Applying the same 
analysis as in the previous scenario, Fig.~\ref{fig:2s2r_photon_entanglement} shows that 
this entanglement, quantified by the negativity defined in 
Eq.~\eqref{eq:def_negativity}, is enhanced even further by the presence of the spins 
as compared to the baseline entanglement of the same cavities in the absence of spins. 
Once again, the spin ensembles act as local `tuning knobs' of the polariton frequencies, 
bringing them into resonance via their interaction with the magnetic field.

The results discussed in this and the previous section illustrate the possibilities 
that this hybrid polariton platform offers for controlling 
spin-spin and photon-photon correlations within a chip by playing with the magnetic field and 
other external parameters. Besides, the margins over which effective spin-spin interactions 
as well as photon entanglement can be varied are determined by the circuit design. Last but not 
least, this theoretical framework provides a solid basis to interpret the microwave transmission 
experiments reported in previous sections. The methods are described in the 
following. 

%%%%%%%%%%%%%%%%%%%%%%%%%%%%%%%%%%%
%%%%%%%%%%%% INPUT-OUTPUT %%%%%%%%%
%%%%%%%%%%%%%%%%%%%%%%%%%%%%%%%%%%%

\subsection{Input-output theory: calculation of microwave transmission}\label{ssec:input-output}

To corroborate the theoretical arguments developed in this section so far, we resort to 
experimental validation. In this work, such validation is provided by microwave 
transmission experiments, in which photons sent through a waveguide interact with the 
system, yielding direct information about its energy level spectrum. To describe these 
experiments, we employ standard input-output theory \cite{gardiner2004quantum}. Starting 
from the Hamiltonian introduced in 
Eqs. (\ref{eq:og_full_hamiltonian}-\ref{eq:res-res-ham}), we map our spins onto harmonic 
oscillators via the Holstein-Primakoff transformation \cite{holstein1940field}. Then, each 
spin ensemble is characterized by its creation and annihilation operators, $c_j^\dagger$ 
and $c_j$, respectively. Solving for the time evolution of the expectation values of 
operators $a_j$, $a_j^\dagger$, $c_j$, and $c_j^\dagger$  by applying standard 
approximations, such as the rotating wave approximation, and focusing on the steady-state 
solution for a perturbative driving of amplitude $\alpha_{\text{in}}$, we arrive at

\begin{widetext}
\begin{equation}
\label{matriz4x4}
%%%%%%%%%%%%%%%%
-i\alpha_{in}
\left(
\begin{array}{c}
\sqrt{\gamma_{c1}} \\
\sqrt{\kappa_{c1}} \\
\sqrt{\kappa_{c2}} \\
\sqrt{\gamma_{c2}}
\end{array}
\right)
=
\underbrace{
\left(
\begin{array}{cccc}
i(\Omega_{S1}-\omega)+\gamma_{1} & iG_{1} & 0 & 0 \\
iG_{1} & i(\omega_{r1}-\omega)+\kappa_{1} & i\kappa_{1,2} & 0 \\
0 & i\kappa_{1,2} & i(\omega_{r2}-\omega)+\kappa_{2} & iG_{2} \\
0 & 0 & iG_{2} & i(\Omega_{S2}-\omega)+\gamma_{2}
\end{array}
\right)
}_{\Mat}
\left(
\begin{array}{c}
\langle c_{1}\rangle \\
\langle a_{1}\rangle \\
\langle a_{2}\rangle \\
\langle c_{2}\rangle 
\end{array}
\right)
%%%%%%%%%%%%%%%%
\;,
\end{equation}
\end{widetext}
where $\kappa_j$ is the overall photon decay rate for $\langle a_j \rangle$ in each LER, 
$\kappa_{c,j}$ the coupling rate of each LER to the readout transmission line, 
$\gamma_j$ the total loss rate for each molecular spin, and $\gamma_{c,j}$ accounts for the 
direct coupling between the spins and the transmission line.

For the present analysis, we neglect the coupling between the spin ensembles and the 
transmission line (i.e., we make $\gamma_{c,j}=0$), as the contact between the deposited 
sample and the line is usually minimal. This simplification enables us to derive an 
analytical expression for the transmission based on the response of each resonator to the 
driving field, which is given by

\begin{equation}
\label{eq:photon_S21}
%%%%%%%%%%%%%%%%
S_{21}
=
1
-
\left(
\sqrt{\kappa_{c1}}
\frac{i\langle a_{1}\rangle}{\alpha_{in}}
+
\sqrt{\kappa_{c2}}
\frac{i\langle a_{2}\rangle}{\alpha_{in}}
\right)
%%%%%%%%%%%%%%%%
\;.
\end{equation}
Now $\langle a_{1}\rangle$ and $\langle a_{2}\rangle$ can be obtained
analytically, by inverting the $4\times4$ matrix on the right-hand
side of Eq.~(\ref{matriz4x4}).
Denoting that matrix by $\Mat$, and applying $\Mat^{-1}$ to both sides
of Eq.~(\ref{matriz4x4}), we have (recall we set $\gamma_{c1}=\gamma_{c2}=0$)
%_____________________________
%
% SOLVING FOR THE AVERAGES
%_____________________________
\begin{widetext}
\begin{eqnarray}
%%%%%%%%%%%%%%%%
\frac{i\langle a_{1}\rangle}{\alpha_{in}}
&=&
\left[
\Mat^{-1}
\left(
\begin{array}{c}
0 \\
\sqrt{\kappa_{c1}} \\
\sqrt{\kappa_{c2}} \\
0
\end{array}
\right)
\right]_{2}
=
(\Mat^{-1})_{2,2}\sqrt{\kappa_{c1}}
+
(\Mat^{-1})_{2,3}\sqrt{\kappa_{c2}}
\label{label29p1}
\\
%%%%%%%%%%%%%%%%
\frac{i\langle a_{2}\rangle}{\alpha_{in}}
&=&
\left[
\Mat^{-1}
\left(
\begin{array}{c}
0 \\
\sqrt{\kappa_{c1}} \\
\sqrt{\kappa_{c2}} \\
0
\end{array}
\right)
\right]_{3}
=
(\Mat^{-1})_{3,2}\sqrt{\kappa_{c1}}
+
(\Mat^{-1})_{3,3}\sqrt{\kappa_{c2}}
\label{label29p2}
%%%%%%%%%%%%%%%%
\;,
\end{eqnarray}
\end{widetext}
%_____________________________
%_____________________________
where we have denoted $\left[\;\cdot\;\right]_{m}$ the $m$th component
of the corresponding column vector.
Thus, we see that we only need four of the matrix elements of
$\Mat^{-1}$ (three actually, as $\Mat$ is symmetric).

In order to get those matrix elements, we write our
3-diagonal symmetric matrix $\Mat$ compactly as follows:
%_____________________________
%
% MATRIX: COMPACT FORM
%_____________________________
\begin{equation}
\label{matrix}
%%%%%%%%%%%%%%%%
\Mat
=
\left(
\begin{array}{cccc}
u_{1} & v_{1} & 0    & 0 \\
v_{1} & u_{2} & v_{2} & 0 \\
0    & v_{2} & u_{3} & v_{3} \\
0    & 0     & v_{3} & u_{4}
\end{array}
\right)
%%%%%%%%%%%%%%%%
\;.
\end{equation}
%_____________________________
%_____________________________
We can get the  required matrix elements of $\Mat^{-1}$ from the
corresponding minors of $\Mat$ by the formula
$(\Mat^{-1})_{j,k}=(-1)^{j+k}M_{k,j}/\det\Mat$.
Here the minor $M_{j,k}$ is the determinant of the submatrix
obtained by deleting the $j$th row and the $k$th column of $\Mat$.

All determinants involved can be written in terms of the following two
combinations
%_____________________________
%
% \theta_{1} & \theta_{2}
%_____________________________
\begin{eqnarray*}
%\label{}
%%%%%%%%%%%%%%%%
\theta_{1}
&=&
u_{1}u_{2}-v_{1}^{2}
\\
%%%%%%%%%%%%%%%%
\theta_{2}
&=&
u_{3}u_{4}-v_{3}^{2}
%%%%%%%%%%%%%%%%
\;,
\end{eqnarray*}
%_____________________________
%_____________________________
which are the determinants of the $2\times2$ diagonal boxes in which
the matrix can be partitioned.
Indeed, $\det\Mat=\theta_{1}\theta_{2}-u_{1}u_{4}v_{2}^{2}$, while
%_____________________________
%
% MATRIX ELEMENTS
%_____________________________
\begin{eqnarray*}
%\label{}
%%%%%%%%%%%%%%%%
(\Mat^{-1})_{2,2}
&=&
u_{1}\theta_{2}/\det\Mat
\\
%%%%%%%%%%%%%%%%
(\Mat^{-1})_{3,3}
&=&
u_{4}\theta_{1}/\det\Mat
\\
%%%%%%%%%%%%%%%%
(\Mat^{-1})_{2,3}
=
(\Mat^{-1})_{3,2}
&=&
-(u_{1}u_{4})v_{2}/\det\Mat
%%%%%%%%%%%%%%%%
\;.
\end{eqnarray*}
%_____________________________
%_____________________________
The last two are duly equal, since $\Mat^{-1}$ ought to be symmetric
too.
The $3\times3$ minors employed come out readily, as the matrix has
many zeroes.
As for $\det\Mat$, it follows by Laplace expansion as:
%_____________________________
%
% BASIC DETERMINANT
%_____________________________
\begin{widetext}
\begin{eqnarray*}
%\label{}
%%%%%%%%%%%%%%%%
\left|
\begin{array}{cccc}
u_{1} & v_{1} & 0    & 0 \\
v_{1} & u_{2} & v_{2} & 0 \\
0    & v_{2} & u_{3} & v_{3} \\
0    & 0     & v_{3} & u_{4}
\end{array}
\right|
&=&
u_{1}
\left|
\begin{array}{ccc}
u_{2} & v_{2} & 0 \\
v_{2} & u_{3} & v_{3} \\
0     & v_{3} & u_{4}
\end{array}
\right|
-
v_{1}
\left|
\begin{array}{ccc}
v_{1} & 0    & 0 \\
v_{2} & u_{3} & v_{3} \\
0     & v_{3} & u_{4}
\end{array}
\right|
\\
%%%%%%%%%%%%%%%%
&=&
u_{1}
\big[
u_{2}
\underbrace{(u_{3}u_{4}-v_{3}^{2})}_{\theta_{2}}
-
u_{4}v_{2}^{2}
\big]
-
v_{1}^{2}
\underbrace{(u_{3}u_{4}-v_{3}^{2})}_{\theta_{2}}
\\
%%%%%%%%%%%%%%%%
&=&
\underbrace{(u_{1}u_{2}-v_{1}^{2})}_{\theta_{1}}
\theta_{2}
-
u_{1}u_{4}v_{2}^{2}
%%%%%%%%%%%%%%%%
\;,
\end{eqnarray*}
\end{widetext}
%_____________________________
%_____________________________
as announced.

Let us now plug the above $\Mat^{-1}$ matrix elements in the
right-hand side of Eqs.~(\ref{label29p1}) and~(\ref{label29p2}) and
then assemble the term in the parenthesis of Eq.~(\ref{eq:photon_S21}).
The result reads
%_____________________________
%
% PARENTHESIS IN EQ.(29)
%_____________________________
\begin{widetext}
\begin{eqnarray*}
%\label{}
%%%%%%%%%%%%%%%%
\sqrt{\kappa_{c1}}
\frac{i\langle a_{1}\rangle}{\alpha_{in}}
+
\sqrt{\kappa_{c2}}
\frac{i\langle a_{2}\rangle}{\alpha_{in}}
=
\frac{
\kappa_{c1}u_{1}\theta_{2}+\kappa_{c2}u_{4}\theta_{1}
-
2\sqrt{\kappa_{c1}\kappa_{c2}}\,
(u_{1}u_{4})v_{2}
}
{\theta_{1}\theta_{2}-(u_{1}u_{4})v_{2}^{2}}
%%%%%%%%%%%%%%%%
\;,
\end{eqnarray*}
\end{widetext}
%_____________________________
%_____________________________
and we are almost done.
We divide numerator and denominator by $u_{1}u_{4}$, introduce
$z_{1}\equiv\theta_{1}/u_{1}=u_{2}-v_{1}^{2}/u_{1}$
and
$z_{2}\equiv\theta_{2}/u_{4}=u_{3}-v_{3}^{2}/u_{4}$,
so we can write
%_____________________________
%
% Eq.(30)
%_____________________________
\begin{equation}
\label{label30}
%%%%%%%%%%%%%%%%
S_{21}
=
1
-
\frac{
\kappa_{c1}z_{2}(\omega)+\kappa_{c2}z_{1}(\omega)
-
2i\sqrt{\kappa_{c1}\kappa_{c2}}\,\kappa_{1,2}
}
{z_{1}(\omega)z_{2}(\omega)+\kappa_{1,2}^{2}}
%%%%%%%%%%%%%%%%
\;.
\end{equation}
%_____________________________
%_____________________________
Here we reverted to the original $-v_{1}^{2}=G_{1}^{2}$ and
$-v_{3}^{2}=G_{2}^{2}$, along with $v_{2}=i\kappa_{1,2}$.
We close with explicit expressions for the important functions $z_{1}$ and
$z_{2}$.
They can be written in an unified way as ($j=1,2$)
%_____________________________
%
% Eq.(31)
%_____________________________
\begin{equation}
\label{eq:z_transmision}
%%%%%%%%%%%%%%%%
z_{j}(\omega)
=
[i(\omega_{rj}-\omega)+\kappa_{j}]
+
\frac{G_{j}^{2}}{[i(\Omega_{Sj}-\omega)+\gamma_{j}]}
%%%%%%%%%%%%%%%%
\;.
\end{equation}
%_____________________________
%_____________________________
The terms in the square brackets are $u_{2}$ \& $u_{1}$, for $j=1$,
and $u_{3}$ \& $u_{4}$, when $j=2$, as given in the original matrix. Equipped with these 
expressions, we can extract the values of the parameters defining our model from the 
experimental measurements.

\begin{figure}[htb!]
\centering
\includegraphics[width=0.95\columnwidth]{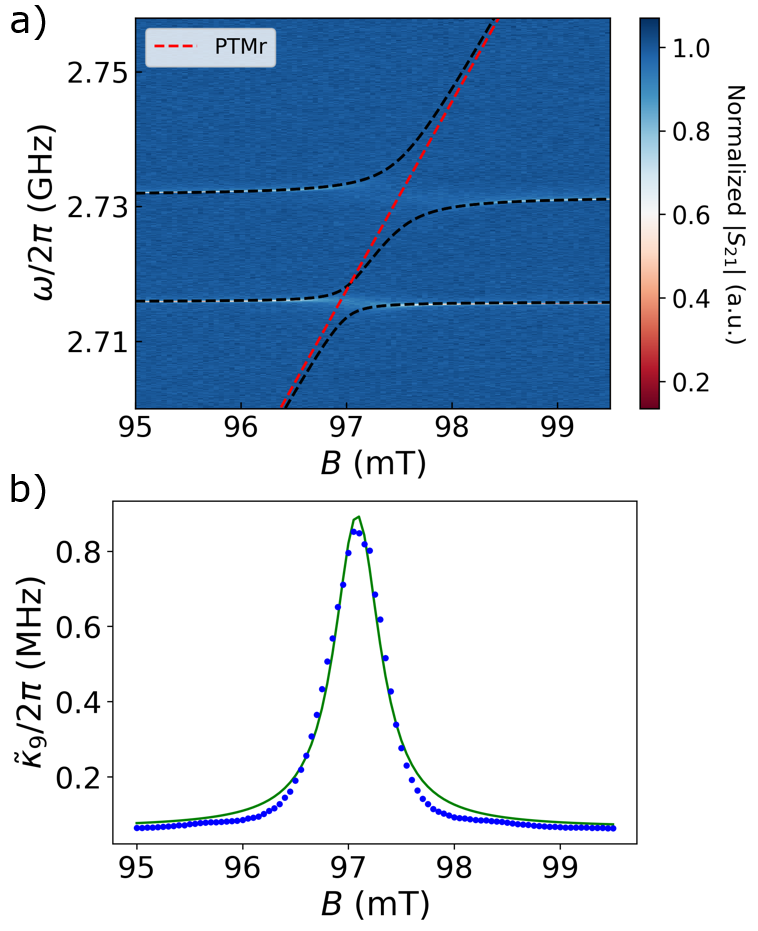}
\caption{\justifying a) Result of fitting the eigenvalues of the Hamiltonian \eqref{eq:spin-normal-modes} to the 
experimental modes determined from microwave transmission data for LER-9 and LER-10 (cf Fig. \ref{fig:remote-spin}). 
LER-9 remains empty while LER-10 hosts a PTMr sample, whose resonance frequency is shown as a red dotted line, 
leading to a local spin-photon collective coupling $G_{10}/2 \pi =5.4$~MHz. b) Effective half width at half maximum 
of LER-9 as a function of magnetic field (dots) and least-squares fitbased on Eq.~\eqref{eq:kappa_weak_coupling}, yielding 
$G_{9,10}/2 \pi = 2.46$~MHz.}
\label{fig:fitting LER9y10}
\end{figure}

\subsection{Application}\label{ssec:application}

\subsubsection{One spin ensemble scenario: microwave transmission and remote spin-photon coupling}\label{sssec:1-spin-application}
Analogous to the theoretical framework developed previously, we begin by examining the case 
where only one LER within a pair couples to a spin ensemble, while the other remains empty. 
This scenario corresponds to the experiments performed with LER-9 and LER-10, shown in 
Fig.~\ref{fig:remote-spin}. Microwave transmission results obtained for this system 
are shown in Fig.~\ref{fig:fitting LER9y10}, which show an 
effective coupling of the spin ensemble to both modes. 

In order to extract the effective 
coupling constants, we adopt the following strategy. First, we treat the 
spin-free LER-9 as if it were a standard LER coupled to a spin ensemble with a 
characteristic coupling constant given by $G_{9,10}$. Using input-output theory in the weak 
coupling regime \cite{Rollano2022}, we fit the magnetic field dependence of 
this LER effective line width $\widetilde{\kappa}_{9}$ with the expression:
\begin{equation}       
\widetilde{\kappa}_{9}=\kappa_{9}+\left[\frac{G_{9,10}^{2}}{(\Omega_{S}-\widetilde{\omega}_{\text{r}})^2+ \gamma^{2}}\right] \gamma\ .
\label{eq:kappa_weak_coupling}
\end{equation} 
The result is shown in Fig.~\ref{fig:fitting LER9y10}b. The fit allows estimating  
$G_{9,10}$, as well as the spin resonance linewidth $\gamma$. 

\begin{figure*}[htb!]
\centering
\includegraphics[width=\linewidth]{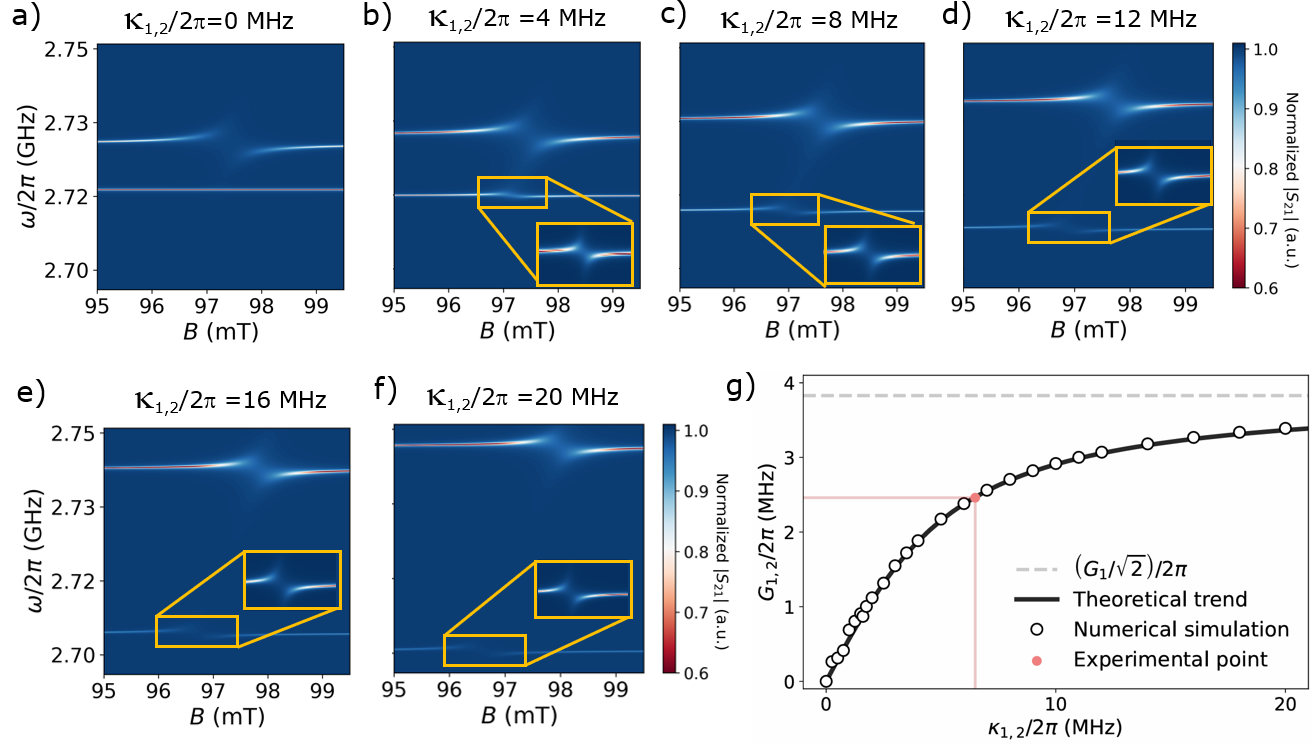}
\caption{\justifying a)-f) $2D$ microwave transmission maps simulated, using Eq.~\eqref{label30}, for 
a coupled LER-1 and LER-2 pair with different values of $\kappa_{1,2}$, as indicated. 
Other relevant parameters are close to those estimated for LER-9 and LER-10, shown in Fig.~\ref{fig:remote-spin}.
The lowest frequency LER-1 remains empty while 
LER-2 couples to a spin ensemble with $G_2/2 \pi =  5.4$~MHz. As $\kappa_{1,2}$ increases, 
the avoided level crossing observed for LER-1 widens. g) Remote coupling of the 
spin ensemble to LER-1, $G_{1,2}$, determined by fitting the LER-1 resonance with  
$|S_{21}|=|1-\frac{\kappa_{\mathrm{c},1}}{z_{1}(\omega)}|$, where 
$z_{1}(\omega)=i(\omega_{\mathrm{r},1}-\omega)+\kappa + \frac{G_{1,2}^2}{i(\Omega_{S,2}-\omega)+\gamma_{2}}$ 
[cf Eq. \eqref{eq:z_transmision}]. The solid line follows the prediction 
given by Eq.~\eqref{eq:g12}. The orange solid dot is the experimental $G_{9,10}$ 
coupling constant, which allows determining $\kappa_{1,2}$.}
\label{fig:G12_k12_LERs9y10}
\end{figure*}

Note that, as discussed in Appendix~\ref{sssec:1-spin-theory} above, the resonance does not occur 
when the polariton frequency 
crosses  $\omega_{\text{r},9}$ but rather $\widetilde{\omega}_-$. Consequently, in 
Eq.~\eqref{eq:kappa_weak_coupling}, $\widetilde{\omega}_{\text{r}} = \widetilde{\omega}_-$. 
Based on the relationship between $\omega_{\text{r},j}$ and $\widetilde{\omega}_{\pm}$ [cf. 
Eq.~\eqref{eq:spin-normal-modes}] along with Eq.~\eqref{eq:g12} for the remote coupling 
$G_{9,10}$, we can determine the bare resonance frequencies of each LER and $\kappa_{1,2}$. 
We find $\kappa_{1,2}/2\pi = 6.49$~MHz, $\omega_{\text{r},1}/2\pi = 2.730$~GHz, 
$\omega_{\text{r},2}/2\pi=2.720$~GHz, $\gamma/2\pi = 7.3$~MHz. The fits are shown in 
Fig.~\ref{fig:fitting LER9y10}. It is worth mentioning that without the presence of the 
spin ensemble in this experiment, determining the cavity parameters 
$\omega_{\rm r,j}$ and $\kappa_{1,2}$ would have been impossible. Since the transmission 
experiment probes the coupled normal modes, we would face an underdetermined system of 
equations, because distinct sets of values for the detuning between bare resonance 
frequencies and the inter-resonator coupling can yield the same level splitting. In this 
scenario, the spin serves to fully characterize the coupled photonic system.

\begin{figure*}[t]
\centering
\includegraphics[width=0.85\linewidth]{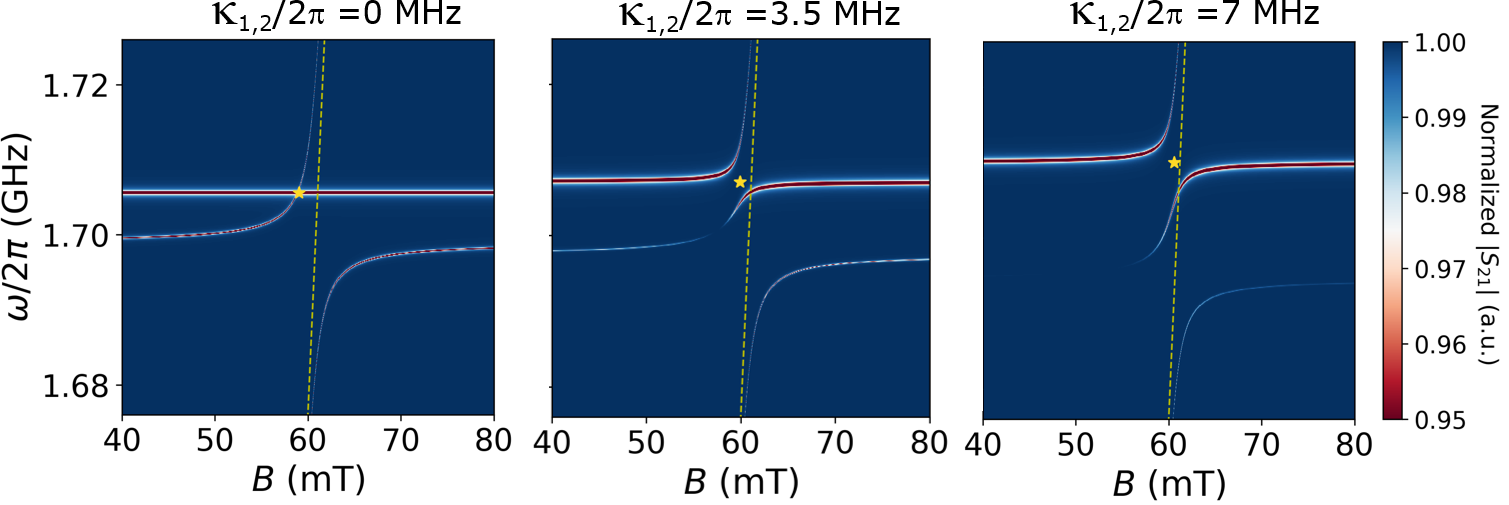}
\caption{\justifying Simulation of the transmission signal for a LER pair coupled to 
one spin ensemble and for different values of $\kappa_{1,2}$. The 
lower frequency LER-1 hosts a molecular sample leading to a local $G_1/2 \pi \approx 20 $~MHz 
and LER-2 is empty. The yellow dotted line denotes the bare $\Omega_{S,1}$ as a function 
of magnetic field. The anti-crossing occurs at the magnetic field where the 
frequency of the polariton mode 
formed at LER-1, not $\Omega_{S,1}$, crosses the frequency of the other normal mode, as predicted by 
\eqref{eq:crossing_field_1s2l} that is marked by a star.}
\label{fig:S21_one_empty}
\end{figure*}

In order to validate this procedure, we perform simulations of the microwave transmission 
with Eq.~\eqref{label30} and with all parameters, save for $\kappa_{1,2}$, as given above.
By varying $\kappa_{1,2}$, we have checked that the remote coupling $G_{9,10}$ derived from 
the fits of microwave transmission based on Eq.~\eqref{eq:kappa_weak_coupling}, follows the 
expected trend, predicted by \eqref{eq:g12}. The results, displayed in 
Fig.~\ref{fig:G12_k12_LERs9y10}, confirm that the analysis of the experimental results 
in Fig.~\ref{fig:fitting LER9y10} do indeed provide a $G_{9,10}$ value consistent with 
theory.

Finally, we also test the prediction of Eq.~\eqref{eq:crossing_field_1s2l} for the crossing 
magnetic field between the polariton formed at the LER that hosts a spin ensemble and the 
normal mode that is closest to the other LER. We consider a LER pair formed by 
LER-1, strongly coupled to a spin ensemble, and the empty LER-2. 
Figure~\ref{fig:S21_one_empty} shows microwave transmission maps 
calculated for different values of $\kappa_{1,2}$. In this figure, a yellow star marks the 
crossing predicted from Eq.~\eqref{eq:crossing_field_1s2l}, which agrees very well with the 
simulation. These simulations also confirm that the crossing approaches the bare spin 
resonance (indicated by a yellow dashed line) as $\kappa_{1,2}$ increases.

\subsubsection{Two spin ensemble scenario: eigenstates, microwave transmission and polariton visibilities}\label{sssec:2-spin-application}

\begin{figure}[h!]
\centering
\includegraphics[width=0.95\columnwidth]{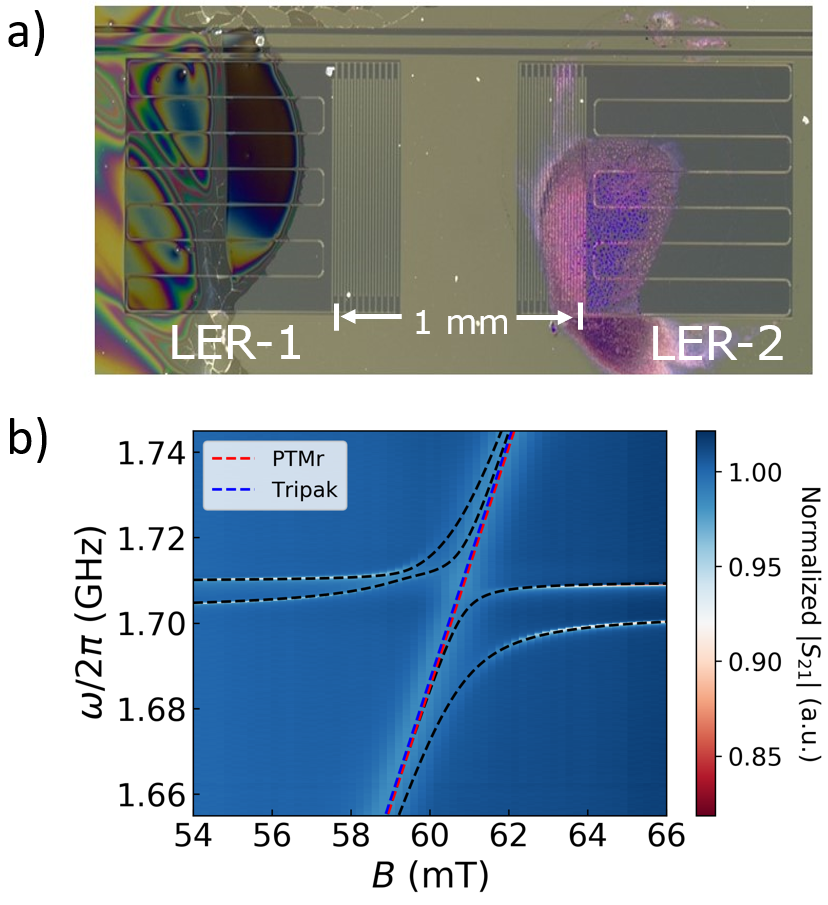}
\caption{\justifying a) Optical microscopy image of the two coupled LER-1 and LER-2, hosting 
$N\sim 5\times 10^{14}$ PTMr and Tripak$^{-}$ free radical molecules, respectively. b) 
$2D$ microwave transmission map measured near their 
resonance frequencies at $T=11$~mK and as a function of magnetic field. The lines 
show the eigenfrequencies obtained by exact diagonalization of the 
Hamiltonian~\eqref{eq:og_full_hamiltonian} using the parameter values derived by fitting 
the experimental transmission with Eq.~\eqref{label30}. We find 
$\omega_{\text{r},1}/2 \pi = 1.703$~GHz, $\omega_{\text{r}2}/2 \pi = 1.710$~GHz, 
$G_1/2 \pi = 19.5$~MHz, $G_2/2 \pi = 8.5$~MHz and $\kappa_{1,2}/2 \pi = 1.06$~MHz.}
\label{fig:autoval_LERs1and2}
\end{figure}

We now turn to the case involving two distinct spin ensembles, each coupled to a separate 
LER. Similarly to the previous case, we can extract the relevant parameters in Eq.~\eqref{eq:og_full_hamiltonian} by fitting 
the experimental transmission data. In this 
way, we are able to determine the local spin-photon couplings ($G_1/2 \pi = 19.5$~MHz, and $G_2/2 \pi = 8.5$~MHz), the inter-
cavity coupling ($\kappa_{1,2}/2 \pi = 1.06$~MHz), and 
the bare LER frequencies ($\omega_{\text{r},1}/2 \pi = 1.7029$~GHz and  
$\omega_{\text{r},2}/2 \pi=1.7096$~GHz). Figure~\ref{fig:autoval_LERs1and2} shows that  the system 
eigenfrequencies obtained by the exact diagonalization of the Hamiltonian
reproduce the experimental modes.

\begin{figure}[h]
\centering
\includegraphics[width=\columnwidth]{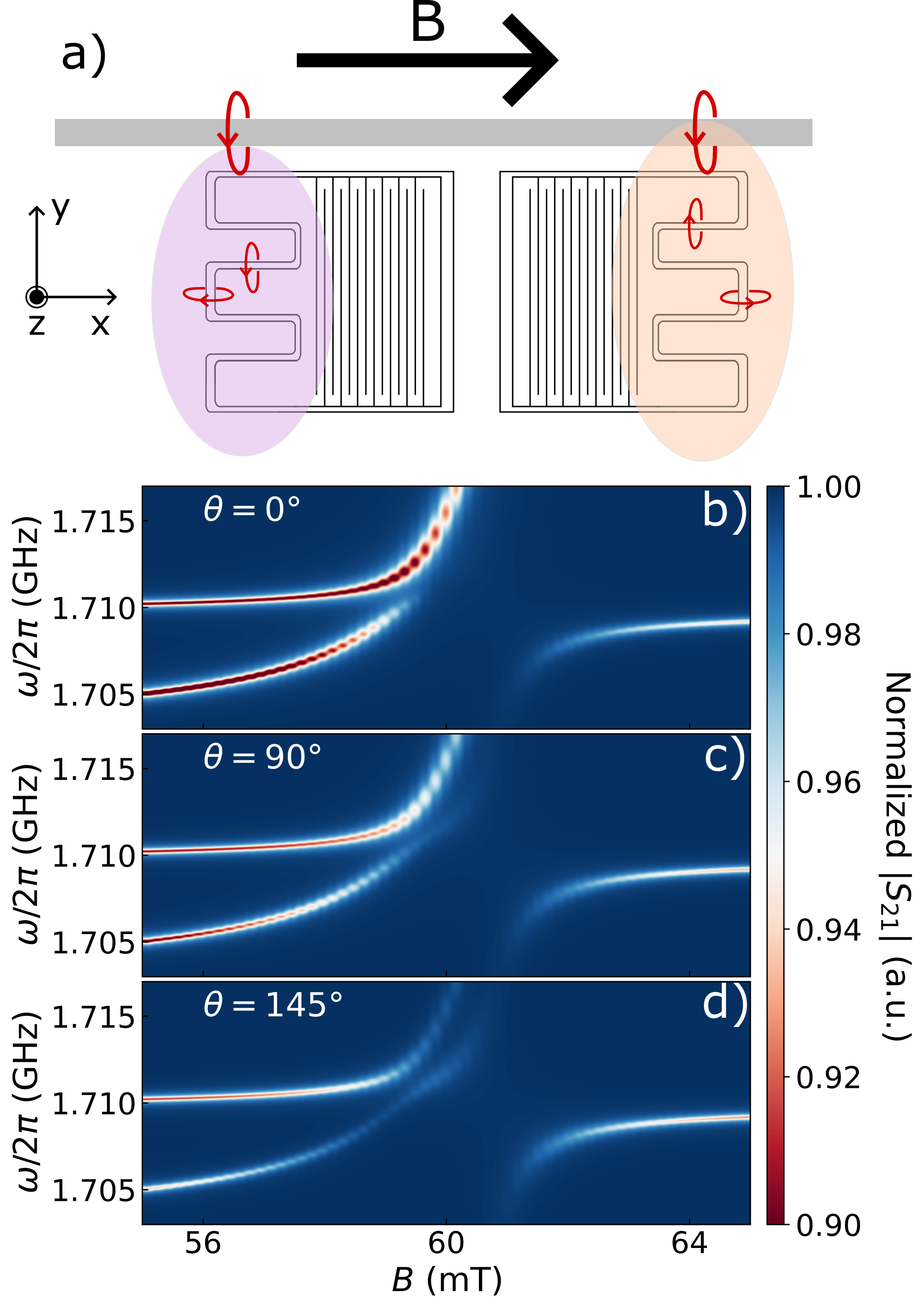}
\caption{\justifying a) Sketch of the LER-1 and LER-2 pair, showing the coupling of the  
Tripak$^{-}$ (purple) and PTMr (orange) molecular spin ensembles to each LER and 
to the readout transmission line (in grey at the top). b-d) Dependence of microwave 
transmission visibility on the effective relative angle between the microwave field 
generated by the transmission line and the field generated by the LER inductor. 
The microwave $2D$ map in panel d) 
corresponds to the final fit shown in Fig.~\ref{fig:model}c}
\label{fig:dark_transmission}
\end{figure}

There is, however, an interesting detail concerning the experimental visibilities of some of 
these modes. Figure~\ref{fig:spin-spin}f shows that one of the two polariton branches becomes 
darker, while the other one becomes brighter near their avoided crossing. Yet, none
of them vanishes. This seems at odds with the theoretical predictions 
derived from Eq.~\eqref{label30} for this system. If one examines the 
wave functions obtained numerically for this model, it is observed that the 
$\vert \psi_3 \rangle$ eigenstate becomes an antisymmetric combination of LER-1 
and LER-2 photons near the anticrossing (Figs.~\ref{fig:2spin2ler_theory}c and \ref{fig:2spin2ler_theory}e). 
According to Eq.~\eqref{eq:photon_S21}, at the point where 
this perfectly antisymmetric superposition occurs null visibility is expected. That is, 
the normalized transmission should reach unity at that point, as we are dealing with a dark 
state. The non-zero visibility observed experimentally calls for some asymmetry in the 
coupling of both LERs to the readout line. 

In order to account for this effect, it is necessary to include 
in the model a direct coupling between the two spin 
ensembles and the readout waveguide, represented in in Eq.~\eqref{matriz4x4} by the 
parameters $\gamma_{c,j}$. Furthermore, given the microscopic geometry of the system (see 
Fig.~\ref{fig:dark_transmission}a), we treat this coupling as a complex number. This 
additional phase accounts for the fact that spins within the ensemble do not all experience 
the same microwave magnetic field, differing not only in magnitude (due to varying 
proximity to the inductor lines) but also in phase (as this particular resonator design 
exhibits non-zero field components in all spatial directions). 
Figure~\ref{fig:dark_transmission} shows that simply treating $\gamma_{cj}$ as a 
non-zero real value is insufficient; rather, the phase $\theta$ modifies the 
visibility of the polariton branch of interest. Given the 
experimental situation, we take $\gamma_{\rm c,1}=\lambda e^{i\theta}$ with 
$\lambda/2\pi = 65$~kHz and $\gamma_{\rm c,2}=0$. This change in Eq.~\eqref{matriz4x4} 
forces to resort to numerical analysis to compute the response for the operators and hence 
the microwave transmission, which now takes the form
\begin{equation}
\label{eq:dark_S21}
S_{21} =  1 - \sqrt{\kappa_{\rm c,1}}\frac{i\langle a_1 \rangle}{\alpha_{\text{in}}} 
- \sqrt{\kappa_{\rm c2}}\frac{i\langle a_2 \rangle}{\alpha_{\text{in}}} 
- \sqrt{\gamma_{\rm c1}}\frac{i\langle c_1\rangle}{\alpha_{\text{in}}}\ .
\end{equation}

By taking into account this effect, we are able to qualitatively reproduce the signal shown 
in Fig.~\ref{fig:spin-spin}. A more refined fit would require accounting for the specific 
field distribution and orientations derived from the microscopic model.

%
%%%%%%%%%%%%%%%%%%%%%%%%%%%%%%%
%% BIBLIOGRAPHY
%%%%%%%%%%%%%%%%%%%%%%%%%%%%%%%
%
\newpage

\bibliographystyle{apsrev4-1}
\bibliography{bibliography}

%\tableofcontents
\end{document}